\newif\ifOneColumn

\ifOneColumn
    \documentclass[journal,12pt,onecolumn,draftclsnofoot,]{IEEEtran}
\else
    \documentclass[journal,twocolumn]{IEEEtran}
\fi

\usepackage{amsmath,graphicx,algorithmic,float,cite,dsfont,amssymb}
\usepackage{amsfonts,multirow,bm,array,setspace} 
\usepackage{textcomp}
\usepackage{subfigure}
\usepackage{psfrag}
\usepackage{color}
\usepackage[linesnumbered,ruled]{algorithm2e}
\usepackage{pifont}
\newcommand{\crossmark}{\ding{55}}

\usepackage[margin=0.87in]{geometry}

\ifCLASSINFOpdf
\else
\fi

\begin{document}

\title{Spatial Deep Learning for Wireless Scheduling}

\author{\IEEEauthorblockN{Wei Cui, \IEEEmembership{Student Member,~IEEE}, Kaiming Shen, \IEEEmembership{Student Member,~IEEE}, and Wei Yu, \IEEEmembership{Fellow,~IEEE}} 
\thanks{Manuscript submitted to IEEE Journal on Selected Areas in
Communications on July 21, 2018, revised \today.
This work is supported by Natural Science and Engineering Research Council
(NSERC) via the Discovery Grant Program and the Canada Research Chairs program.  The materials in this paper
have been be presented in part at the IEEE Global Communications Conference
(Globecom), Abu Dhabi, December 2018. The authors are with
The Edward S.~Rogers Sr.~Department of Electrical and Computer
Engineering, University of Toronto, Toronto, ON M5S 3G4, Canada
(e-mails: \{cuiwei2, kshen, weiyu\}@ece.utoronto.ca).}%
}


\maketitle


\begin{abstract}
The optimal scheduling of interfering links in a dense wireless network with
full frequency reuse is a challenging task. The traditional method involves
first estimating all the interfering channel strengths then optimizing the
scheduling based on the model. This model-based method is however resource
intensive and computationally hard because channel estimation is expensive 
in dense networks; furthermore, finding even a locally optimal solution of the
resulting optimization problem may be computationally complex. This paper shows
that by using a deep learning approach, it is possible to bypass the channel
estimation and to schedule links efficiently based solely on the geographic
locations of the transmitters and the receivers, due to the fact that in many propagation environments, the wireless channel strength is
largely a function of the distance dependent path-loss. This is accomplished 
by unsupervised training over randomly deployed networks, and 
by using a novel neural network
architecture that computes the geographic spatial convolutions of the interfering
or interfered neighboring nodes along with subsequent multiple feedback stages to learn
the optimum solution. The resulting neural network gives near-optimal
performance for sum-rate maximization and is capable of generalizing to larger
deployment areas and to deployments of different link densities. Moreover, to
provide fairness, this paper proposes a novel scheduling approach that utilizes
the sum-rate optimal scheduling algorithm over judiciously chosen subsets of
links for maximizing a proportional fairness objective over the network. The
proposed approach shows highly competitive and generalizable network utility
maximization results.
\end{abstract}


\IEEEpeerreviewmaketitle


\section{Introduction}

Scheduling of interfering links is one of the most fundamental tasks in
wireless networking. Consider a densely deployed device-to-device
(D2D) network with full frequency reuse, in which nearby links produce
significant interference for each other whenever they are simultaneously
activated. The task of scheduling amounts to judiciously activating
a \emph{subset} of mutually ``compatible'' links so as to avoid excessive
interference for maximizing a network utility.

The traditional approach to link scheduling is based on the paradigm of first
estimating the interfering channels (or at least the interference graph
topology), then optimizing the schedule based on the estimated channels.
This model-based approach, however, suffers from two key shortcomings.
First, the need to estimate not only the direct channels but also all the
interfering channels is resource intensive.  In a network of $N$
transmitter-receiver pairs, $N^2$ channels need to be estimated within each
coherence block. Training takes valuable resources away from the actual data
transmissions; further, pilot contamination is inevitable in large networks.
Second, the achievable data rates in an interfering environment are
nonconvex functions of the transmit powers. Moreover, scheduling variables
are binary. Hence, even with full channel knowledge, the optimization
of scheduling is a nonconvex integer programming problem for which finding
an optimal solution is computationally complex and is challenging for
real-time implementation.

This paper proposes a new approach, named \emph{spatial deep learning}, to address
the above two issues. Our key idea is to recognize that in many deployment
scenarios, the optimal link scheduling does not necessarily require the exact
channel estimates, and further the interference pattern in a network is to a
large extent determined by the relative locations of the transmitters and
receivers. Hence, it ought to be possible to \emph{learn} the optimal
scheduling based solely on the geographical locations of the neighboring
transmitters/receivers, thus bypassing channel estimation altogether. Toward
this end, this paper proposes a neural network architecture that computes the
geographic spatial convolution of the interfering or interfered neighboring
transmitters/receivers and learns the optimal scheduling in a densely
deployed D2D network over multiple stages based on the spatial parameters alone.

We are inspired by the recent explosion of successful applications of machine
learning techniques \cite{lecun_gradient,lecun_deep} that demonstrate the
ability of deep neural networks to learn rich patterns and to approximate
arbitrary function mappings \cite{hornik}.  We further take advantage of the
recent progress on fractional programming methods for link scheduling
\cite{shen_ISIT17,part1,part2} that allows us to compare against the
state-of-the-art benchmark. The main
contribution of this paper is a specifically designed neural network
architecture that facilitates the spatial learning of geographical locations of
interfering or interfered nodes and is capable of achieving a large portion of
the optimum sum rate of the state-of-the-art algorithm in a computationally
efficient manner, while requiring no explicit channel state information (CSI).

Traditional approach to scheduling over wireless interfering links
for sum rate maximization are all based on (non-convex) optimization, e.g.,
greedy heuristic search \cite{FlashLinQ}, iterative methods for achieving
quality local optimum \cite{shen_ISIT17, luo_TSP11}, methods based on
information theory considerations \cite{ITLinQ,ITLinQ+} or hyper-graph coloring
\cite{Guo_TCOM17,color}, or methods for achieving the global optimum but with
worst-case
exponential complexity such as polyblock-based optimization \cite{MAPEL} or
nonlinear column generation \cite{Johansson_TWC06}. The recent re-emergence of machine learning has motivated the use of neural networks for wireless network optimization. This paper
is most closely related to the recent work of \cite{hong_spawc, alejandro} in adapting
deep learning to perform power control and \cite{cong_ensemble} in utilizing
ensemble learning to solve a closely related problem, but we go one step further than
\cite{hong_spawc, cong_ensemble, alejandro} in that we forgo the traditional
requirement of CSI for spectrum optimization. We demonstrate that for wireless networks in which the channel gains largely depend on the
path-losses, the location information (which can be easily obtained via global
positioning system) can be effectively used as a proxy for obtaining
near-optimum solution, thus opening the door for much wider application of
learning theory to resource allocation problems in wireless networking.

The rest of the paper is organized as follows. Section II establishes the
system model. Section III proposes a deep learning based approach
for wireless link scheduling for sum-rate maximization.
The performance of the proposed method is provided in Section IV. Section V
discusses how to adapt the proposed method for proportionally fair scheduling.
Conclusions are drawn in Section VI.

\section{Wireless Link Scheduling}

Consider a scenario of $N$ independent D2D links located in a two-dimensional
region. The transmitter-receiver distance can vary from
links to links. We use $p_i$ to denote the fixed transmit power level of the
$i$th link, if it is activated. Moreover, we use $h_{ij}\in\mathbb C$ to denote
the channel from the transmitter of the $j$th link to the receiver of the $i$th
link, and use $\sigma^2$ to denote the background noise power level. Scheduling
occurs in a time slotted fashion. In each time slot, let
$x_i\in\{0,1\}$ be an indicator variable for each link $i$, which equals to 1
if the link is scheduled and 0 otherwise. We assume full frequency reuse with
bandwidth $W$. Given a set of scheduling decisions $x_i$, the achievable rate
$R_i$ for link $i$ in the time slot can be computed as
\begin{align} \label{equ:instantRate}
R_i =W\log\left(1+\frac{|h_{ii}|^2p_ix_i}{\Gamma(\sum_{j\neq i}|h_{ij}|^2 p_jx_j + \sigma^2)}\right),
\end{align}
where $\Gamma$ is the signal-to-noise ratio (SNR) gap to the information theoretical channel capacity, due to the use of practical coding and modulation for the linear Gaussian channel \cite{forney}.
Because of the interference between the links, activating all the links at the
same time would yield poor data rates. The wireless link scheduling problem is
that of selecting a subset of links to activate in any given transmission period
so as to maximize some network utility function of the achieved rates.

This paper considers the objective function of maximizing the weighted sum rate
over the $N$ users over each scheduling slot. More specifically, for fixed
values of weights $w_i$, the scheduling problem is formulated as
\begin{subequations}
\label{prob}
\begin{align}
\underset{\mathbf{x}}{\text{maximize}}\quad&
\sum^N_{i=1} w_i R_i\\
\text{subject to}\quad& x_i \in \{0,1\},\;\forall i.
\end{align}
\end{subequations}
The weights $w_i$ indicate the priorities assigned to each user, (i.e., the higher
priority users are more likely to be scheduled). The overall problem is
a challenging discrete optimization problem, due to the complicated
interactions between different links through the interference terms in the
signal-to-interference-and-noise (SINR) expressions, and the different
priority weights each user may have.

The paper begins by treating the scheduling problem with equal weights $w_1 = w_2 =
\cdots = w_N$, equivalent to a sum-rate maximization problem. The second part
of this paper deals with the more challenging problem of scheduling under
adaptive weights $w_1, w_2,\cdots,w_N$ for maximizing a network utility. The assignment of weights is typically based on upper-layer considerations, e.g., as function
of the queue length in order to minimize delay or to stabilize the queues
\cite{neely}, or as function of the long-term average rate of each user in
order to provide fairness across the network \cite{berry}, or as combination of both.

This paper utilizes unsupervised training to optimize the parameters of the
neural network. The results will be compared against multiple benchmarks
including a recently developed and the-state-of-art fractional programming 
approach (referred to as FPLinQ or FP) \cite{shen_ISIT17} for obtaining high-quality 
local optimum benchmark solutions. We remark that the FPLinQ benchmark
solutions can also be utilized as training targets for supervised training 
of the neural network, and 
a numerical comparison is provided later in the paper. FPLinQ relies on a transformation of the SINR expression
that decouples the signal and the interference terms and a subsequent
coordinated ascent approach to find the optimal transmit power for all the
links.  The FPLinQ algorithm is closely related to the weighted minimum
mean-square-error (WMMSE) algorithm for weighted sum-rate maximization
\cite{luo_TSP11}. For the scheduling task, FPLinQ quantizes the optimized power
in a specific manner to obtain the optimized binary scheduling variables.

\section{Deep Learning Based Link Scheduling for Sum-Rate Maximization}

We begin by exploring the use of deep neural network for scheduling, while
utilizing only location information, under the sum-rate maximization criterion.
The sum-rate maximization problem (i.e., with equal weights) is considerably
simpler than weighted rate-sum maximization, because all the links have equal
priority.  We aim to use path-losses and the geographical locations information 
to determine which subset of links should be scheduled.

\begin{figure*}
        \centering
        \ifOneColumn
\else
            \centerline{\includegraphics[width=10.5cm]{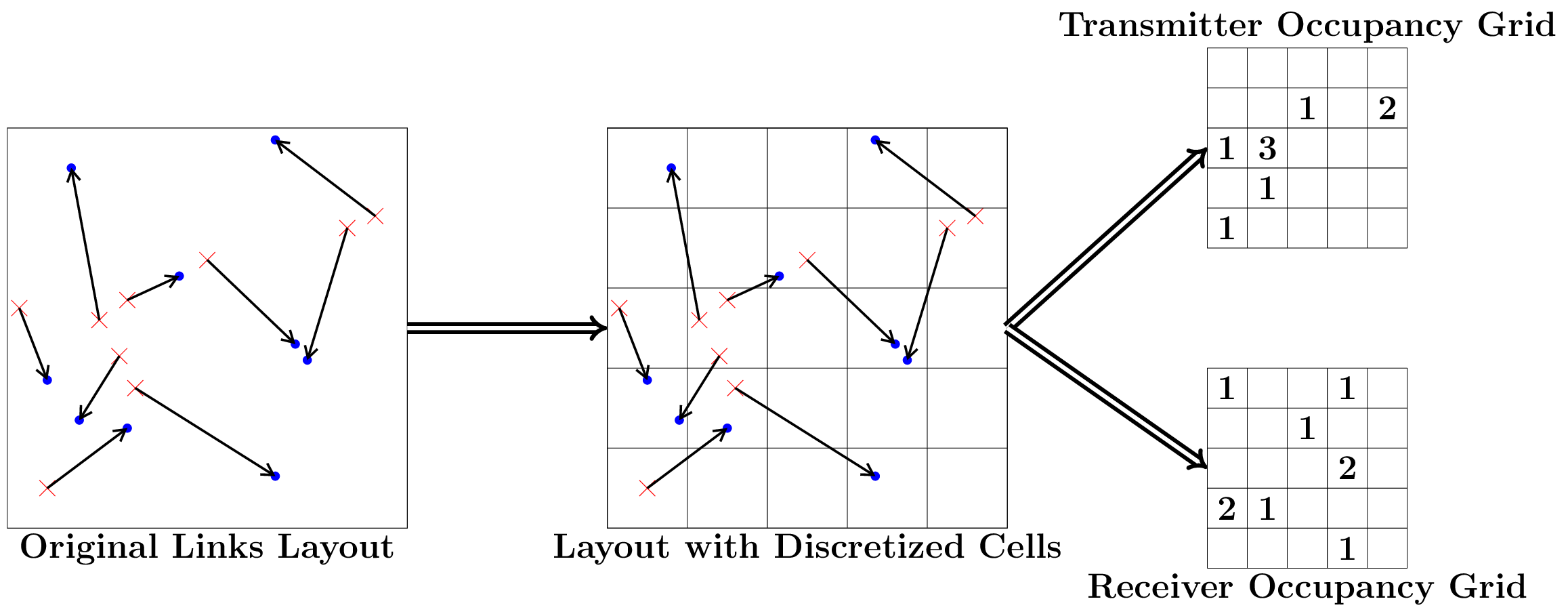}}
        \fi
        \caption{Transmitter and receiver density grids}
        \label{fig:gridexplain}
\end{figure*}

\subsection{Learning Based on Geographic Location Information}

A central goal of this paper is to demonstrate that for wireless networks 
in which the channel gains are largely functions of distance dependent path-losses,
the geographical location information is already sufficient as a proxy for
optimizing link scheduling.  This is in contrast to traditional optimization
approaches for solving (\ref{prob}) that require the full instantaneous CSI, 
and also in contrast to the recent work \cite{hong_spawc}
that proposes to use deep learning to solve the power control problem by
learning the WMMSE optimization process. In \cite{hong_spawc}, a fully
connected neural network is designed that
takes in the channel coefficient matrix as the input, and produces optimized
continuous power variables as the output to maximize the sum rate.
While satisfactory scheduling performance has been obtained in \cite{hong_spawc},
the architecture of \cite{hong_spawc} is not scalable. In a D2D links network
with $N$ transmitter-receiver pairs, there are $N^2$ channel coefficients. A
fully connected neural network with $N^2$ nodes in the input layer and $N$
output layer would require at least $O(N^3)$ interconnect weights (and most
likely much more). Thus, the neural network architecture proposed in
\cite{hong_spawc} has training and testing complexity that grows rapidly with
the number of links.

Instead of requiring the full set of CSI between every
transmitter and every receiver as the inputs to the neural network $\{h_{ij}\}$,
which has $O(N^2)$ entries, this paper proposes to use the geographic location information (GLI) as input,
defined as a set of vectors $\{(\mathbf d^{\text{tx}}_{i},\mathbf
d^{\text{rx}}_{i})\}_i$, where $\mathbf d^{\text{tx}}_{i}\in\mathbb R^2$ and
$\mathbf d^{\text{rx}}_{i}\in\mathbb R^2$ are the transmitter and the receiver
locations of the $i$th link, respectively. Note that the input now
scales linearly with the number of links, i.e., $O(N)$.

We advocate using GLI as a substitute for CSI because in many wireless
deployment scenarios, GLI already captures the main feature of channels: the
path-loss and shadowing of a wireless link are mostly functions of distance and
location. This is essentially true for outdoor wireless channels, and
especially so in rural regions or remote areas, where the number of
surrounding objects to reflect the wireless signals is sparse. An example
application is a sensor network deployed outdoors for environmental monitoring
purposes.   

In fact, if we account for fast fading in addition, the CSI can be thought of
as a stochastic function of GLI 
\begin{equation}
\text{CSI} = f(\text{GLI}).
\end{equation}
While optimization approaches to the wireless link scheduling problem aim to
find a mapping $g(\cdot)$ from CSI to the scheduling decisions, i.e.,
\begin{equation}
\mathbf x = g(\text{CSI}),
\end{equation}
the deep learning architecture of this paper aims to capture directly the
mapping from GLI to $\mathbf x$, i.e., to learn the function
\begin{equation}
\mathbf x = g(f(\text{GLI})).
\end{equation}

\subsection{Transmitter and Receiver Density Grid as Input}

To construct the input to the neural network based on GLI, we quantize the continuous
$(\mathbf d^{\text{tx}}_{i}, \mathbf d^{\text{rx}}_{i})$ in a grid form.
Without loss of generality, we assume a square $\ell \times \ell$ meters deployment area, partitioned
into equal-size square cells with an edge length of $\ell/M$, so that
there are $M^2$ cells in total. We use $(s,t)\in[1:M]\times[1:M]$ to index the
cells. For a particular link $i$, let $(s^\text{tx}_i,t^\text{tx}_i)$ be the
index of the cell where the transmitter $\mathbf d^\text{tx}_i$ is located, and
$(s^\text{rx}_i,t^\text{rx}_i)$ be the index of the cell where the receiver
$\mathbf d^\text{rx}_i$ is located.  We use the tuple
$(s^\text{tx}_i,t^\text{tx}_i,s^\text{rx}_i,t^\text{rx}_i)$ to represent the
location information of the link.

We propose to construct two \emph{density grid} matrices of size $M \times
M$, denoted by $T$ and $R$, to represent the density of the \emph{active}
transmitters and receivers, respectively, in the geographical area. The density
grid matrices are constructed by simply counting the total number of active
transmitters and receivers in each cell, as illustrated in Fig.~\ref{fig:gridexplain}.
The activation pattern $\{ x_i \}$ is initialized as a vector of all 1's at the
beginning. As the algorithm progressively updates the activation pattern, the
density grid matrices are updated as
\begin{eqnarray}
\label{eq:grid_tx}
T(s,t) & = & \sum_{\{i | (s^\text{tx}_i,t^\text{tx}_i) = (s,t) \}} x_i, \\
R(s,t) & = & \sum_{\{i | (s^\text{rx}_i,t^\text{rx}_i) = (s,t) \}} x_i.
\label{eq:grid_rx}
\end{eqnarray}

\subsection{Novel Deep Neural Network Structure}

The overall neural network structure for link scheduling with sum-rate
objective is an iterative
computation graph. A key novel feature of the network structure is a forward
path including two stages: a convolution stage that captures the interference
patterns of neighboring links based on the geographic location information and
a fully connected stage that captures the nonlinear functional mapping of the
optimized schedule. Further, we propose a novel \emph{feedback connection} between
the iterations to update the state of optimization. The individual stages and the overall network structure are described in detail
below.

\subsubsection{Convolution Stage}

\begin{figure}
        \centering
        \includegraphics[width=7.5cm]{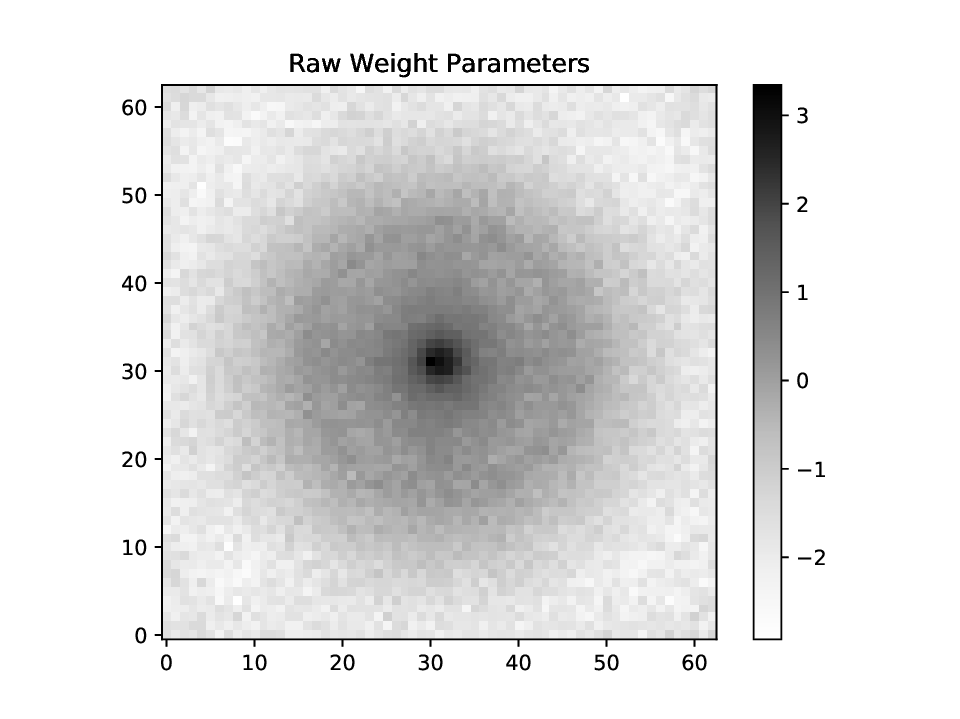}
        \caption{A trained spatial convolution filter (in log scale)}
        \label{fig:convfilter}
\end{figure}

The convolution stage is responsible for computing two functions, corresponding
to that of the interference each link causes to its neighbors and the
interference each link receives from its neighbors, respectively. As a main
innovation in the neural network architecture, we propose to use spatial convolutional
filters, whose coefficients are optimized in the training process, that
operate directly on the transmitter and receiver density grids described in
the previous section. The transmitter and receiver spatial convolutions are
computed in parallel on the two grids. At the end, two pieces of information
are computed for the transmitter-receiver pair of each link: a convolution of spatial
geographic locations of all the nearby receivers that the transmitter can
\emph{cause interference to}, and a convolution of spatial geographic locations of all
the nearby transmitters that the receiver can \emph{experience interference from}.
The computed convolutions are referred to as TxINT$_i$ and RxINT$_i$, respectively,
for link $i$.

Since the idea is to estimate the effect of total interference each link causes
to nearby receivers and effect of the total interference each link is exposed
to, we need to exclude the link's own transmitter and receiver in computing the
convolutions. This is done by subtracting the contributions each link's
own transmitter and receiver in the respective convolution sum.

The convolution filter is a 2D square matrix with fixed pre-defined size and
trainable parameters. The value of each entry of the filter can be interpreted
as the channel coefficient of a transceiver located at a specific distance from
the center of the filter. Through training, the filter
learns the channel coefficient by adjusting its weights.
Fig.~\ref{fig:convfilter} shows a trained filter. As expected, the trained filter
exhibits a circular symmetric pattern with radial decay.

\begin{figure}
        \centering
        \ifOneColumn
\else
            \includegraphics[width=8cm]{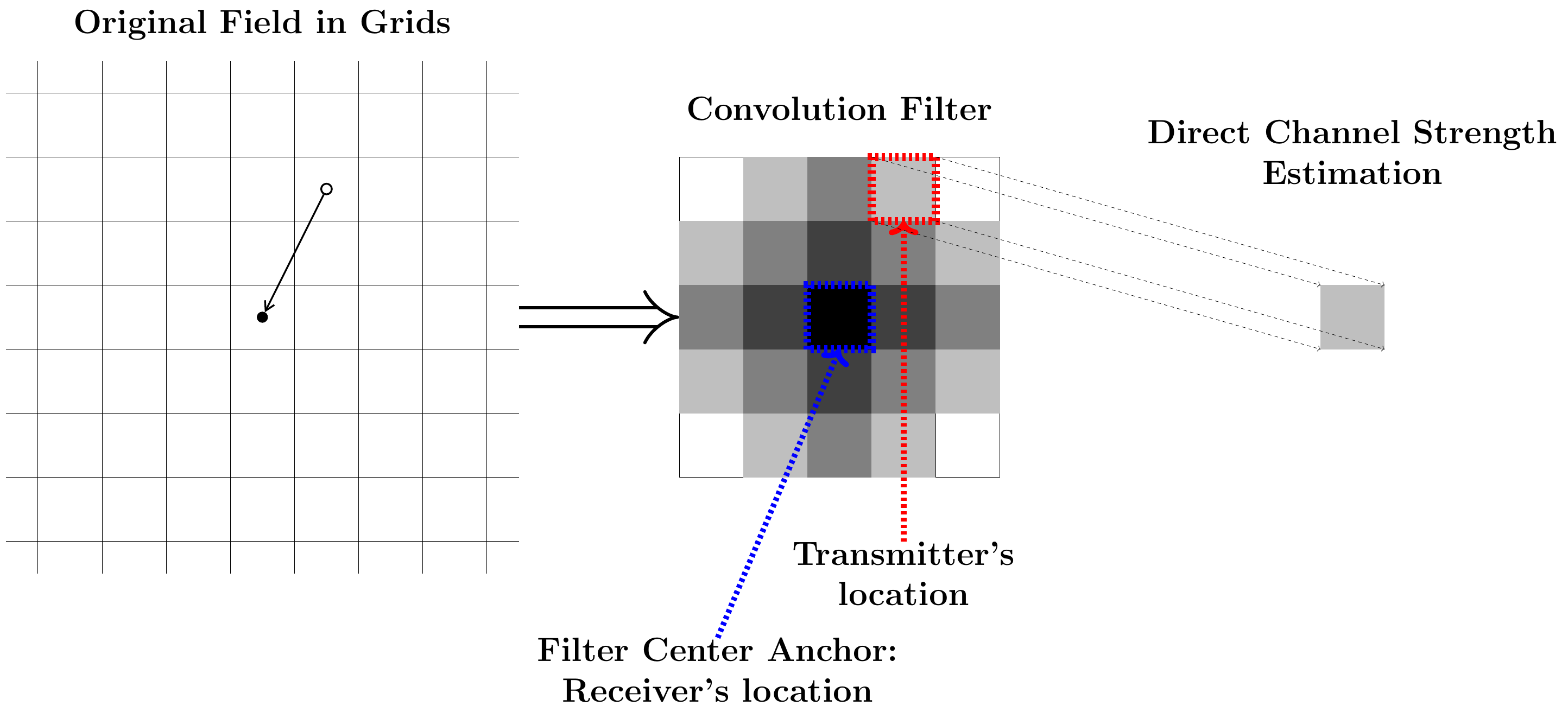}
        \fi
        \caption{Extracting direct channel strength from convolution filter}
        \label{fig:directchannelstrength}
\end{figure}

The convolution stage described above summarizes two quantities for each link:
the total interference produced by the transmitter and the total interference
the receiver is being exposed to. 
Furthermore, we can also extract another important quantity for scheduling from
the trained convolution filter: the \emph{direct channel strength}. At the
corresponding relative location of the transmitter from its receiver, the value
of the convolution filter describes the channel gain of the direct link between this transmitter/receiver pair. The procedure for obtaining this direct channel strength is illustrated in Fig.~\ref{fig:directchannelstrength}. The direct channel strength is referred to as DCS$_i$ for link $i$.

\subsubsection{Fully Connected Stage}
The fully connected stage is the second stage of the forward computation path,
following the convolution stage described above. It takes a feature vector
extracted for each link as input and produces an output $x_i \in [0,1]$,
(which can be interpreted as a relaxed scheduling variable or alternatively as
continuous power) for that link.

\begin{figure*}
        \centering
        \ifOneColumn
\else
            \centerline{\includegraphics[width=10.5cm]{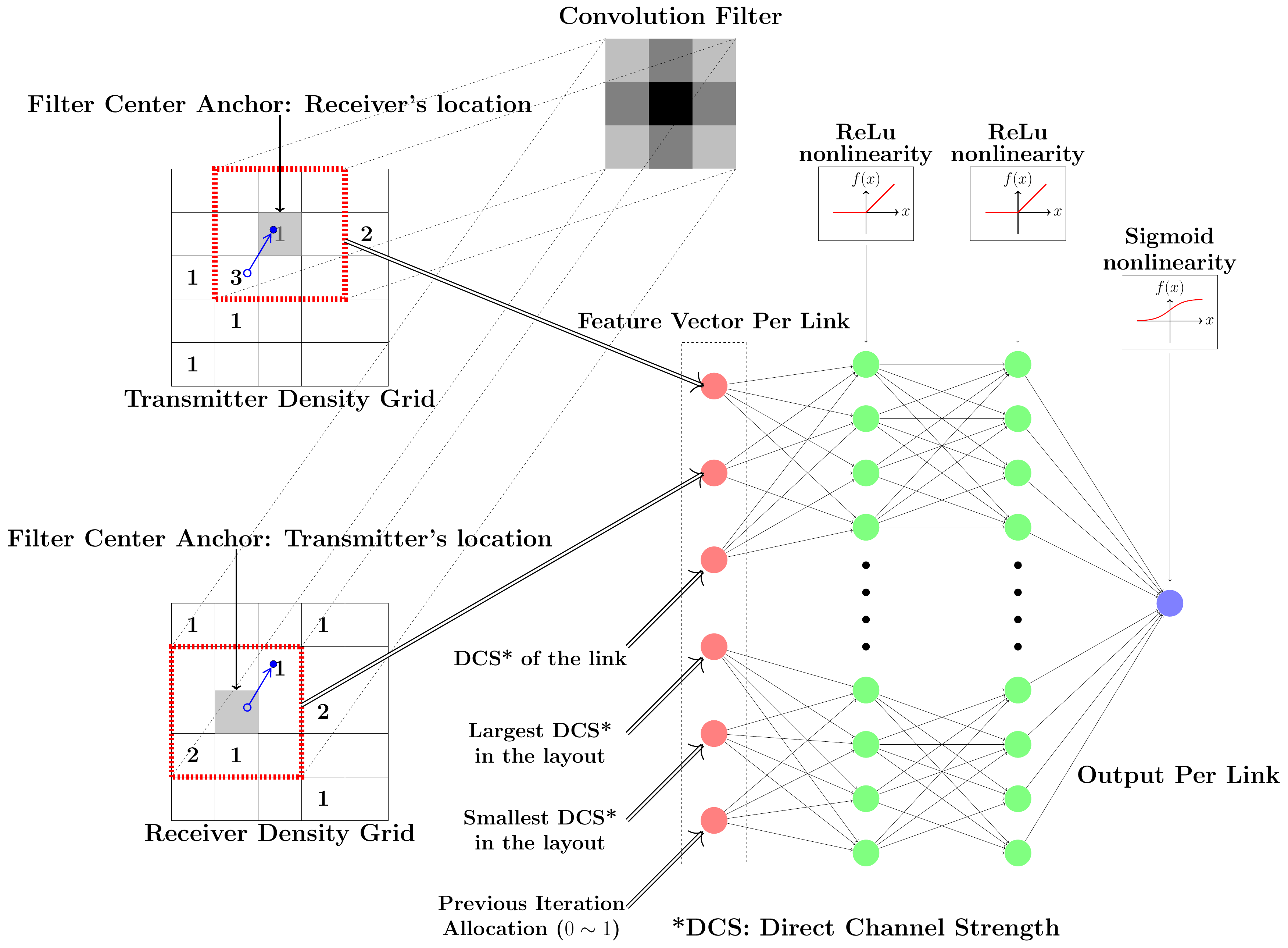}}
        \fi
        \caption{Forward computation path for a single link with spatial convolutions and link distance as input to a neural network}
        \label{fig:fpath}
\end{figure*}

The feature vector for each link comprises of the following entries: TxINT$_i$,
RxINT$_i$, DCS$_i$, DCS$_{\max}$, DCS$_{\min}$, $x_{i}^{t-1}$. The first three
terms have been explained in the previous section. DCS$_{\max}$ and DCS$_{\min}$
denote the largest and smallest direct channel strength among links in the
entire layout; and $x_{i}^{t-1}$ represents the fully connected stage output at
the previous iteration in the overall feedback structure, as described later.
The tuple (TxINT$_i$, RxINT$_i$) describes the interference between
the $i$th link and its neighbors, while the triplet (DCS$_i$, DCS$_{\max}$,
DCS$_{\min}$) describes the link's own channel strength as compared to the 
strongest and the weakest links in the entire layout. 

It is worth noting that the minimum and maximum channel strengths over the
layout are chosen here to characterize the range of the direct channel 
strengths. This is appropriate when the D2D link pairwise distances are roughly
uniform, as we assume in the numerical simulations of this paper. However, if
the D2D link pairwise distances do not follow a uniform distributions, a more
robust characterization could be, for example, $10${th} and $90${th} percentile
values of the channel strength distribution, to alleviate the effect of
potential outliers.  

The value $x_i$ for this link is computed based on its feature vector through
the functional mapping of a fully connected neural network (denoted here as $F_{fc}$)
over the feedback iterations indexed by $t$: 

\ifOneColumn
\else
\begin{multline}
    x_{i}^{t} \gets F_{fc}(\text{TxINT}_{i}, \text{RxINT}_{i}, \text{DCS}_{i}, \\
	\text{DCS}_{\max}, \text{DCS}_{\min}, x_{i}^{t-1}).
\end{multline}
\fi

The convolution stage and the fully connected stage together form one forward
computation path for each transmitter-receiver pair, as depicted 
in Fig.~\ref{fig:fpath}. In the implementation, we use two hidden layers with 30 neurons in each layer to ensure
sufficient expressive power of the neural network. A rectified linear unit
(ReLU) is used at each neuron in the hidden layers; a sigmoid nonlinearity is
used at the output node to produce a value in $[0,1]$.

\subsubsection{Feedback Connection}\label{sec:feedbacksection}

The forward computation (which includes the convolution stage and the fully
connected stage) takes the link activation pattern $x_i$ as the input for
constructing the density grid. In order to account for the progressive
(de)activation pattern of the wireless links through the iterations (i.e., each
subsequent interference estimates need to be aware of the fact that the
deactivated links no longer produce or are subject to interference), we
propose a feedback structure, in which each iteration of the neural network
takes the continuous output $x$ from the previous iteration as input, then
iterate for a fixed number of iterations. We find experimentally that the
network is then able to converge within a small number of iterations. 

The feedback stage is designed as following: After the completion of $(t-1)$th
forward computation, the $\mathbf x$ vector of $[0,1]$ values is obtained, with
each entry representing the activation status for each of the $N$ links. Then,
a new forward computation is started with input density grids prepared by
feeding this $\mathbf x$ vector into (\ref{eq:grid_tx})-(\ref{eq:grid_rx}).
In this way, the activation status for all $N$ links are updated in the density
grids for subsequent interference estimations. 
Note that the trainable weights of the convolutional filter and the neural
network are tied together over the multiple iterations for more efficient training.

\begin{figure*}[t!]
        \centering
        \ifOneColumn
\else
            \includegraphics[width=9cm]{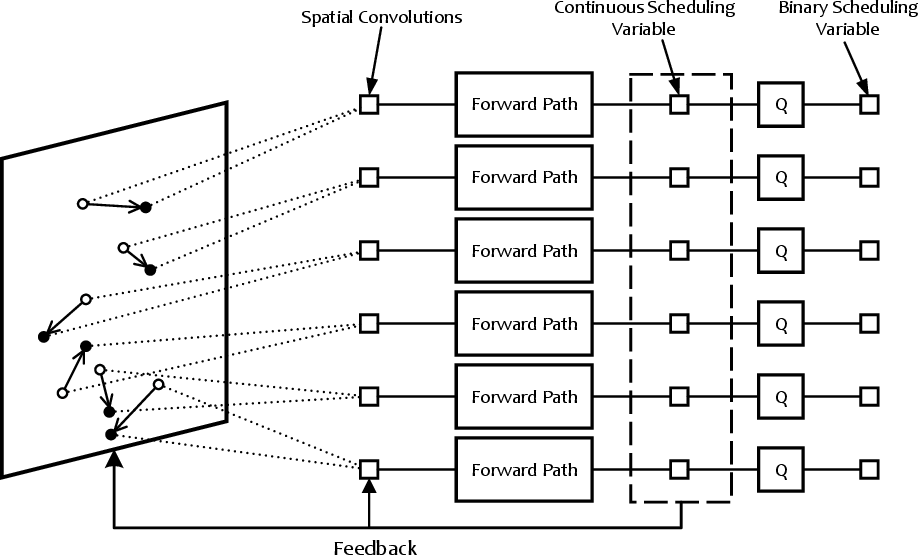}
        \fi
        \caption{Overall neural network with one forward path per link and with
feedback connections and quantized output (denoted as ``Q'').}
        \label{fig:feedback}
\end{figure*}

After a fixed number of iterations, the scheduling decisions are obtained
from the neural network by quantizing the $\mathbf x$ vector from the
last iteration into binary values, representing the scheduling decisions of
the $N$ links.

The overall feedback structure is depicted in Fig.~\ref{fig:feedback}.
We emphasize here that the neural network is designed on a per-link basis, thus
the overall model is scalable with respect to the network size.  Specifically,
at the convolution stage, the convolutions are computed based on the fixed (and
trained) convolution filter that covers the neighboring non-negligible
interference sources. At the fully connected stage, the neural networks of different
links are decoupled, thus scheduling can be performed in a distributed fashion. 

Moreover, in the training stage, the convolutional filter parameters and the
neural network weights of the different links are tied together. This facilitates 
efficient training, and implicitly assumes that the propagation environments of
the different links are similar. Under this homogeneity assumption, 
regardless of how large the layout is and how many links are to be scheduled 
in the network, the overall trained neural network model can be directly
utilized for scheduling, without adjustment or re-training,

\subsection{Training Process}

The overall deep neural network is trained using wireless network layouts with
randomly located links and with the transmitter-receiver distances following a
specific distribution. Specifically, we train the model to maximize the target
sum rate via gradient descent on the convolutional filter weights and neural network
weight parameters. It is worth noting that while the channel gains are needed at the training stage
for computing rates, they are not needed for scheduling, which only requires GLI.

To allow the gradients to be back-propagated through the network, we do not
discretize the network outputs when computing the rates. Therefore, the
unsupervised training process is essentially performing a power control task
for maximizing the sum rate. The scheduling decisions are obtained from
discretizing the optimized power variables. 

We randomly generate wireless D2D networks consisting of $N=50$ D2D pairs in a
500 meters by 500 meters region. The locations for the transmitters are
generated uniformly within the region. The locations of the receivers are
generated according to a uniform distribution within a pairwise distances of
$d_{\min}\sim d_{\max}$ meters from their respective transmitters. We generate
800,000 such network layouts for training.

The transmitter-receiver distance has significant effect on the achievable
rate. Link scheduling for sum-rate maximization tends to favor short links over
long links, 
so the distribution of the link distances has significant effect on the
scheduling performance.
To develop the capacity of the proposed deep learning approach to accommodate 
varying transmitter-receiver distances, we
generate training samples based on the following distribution: 
\begin{itemize}
\item Generate $d_{\min}$ uniformly from $2\sim70$ meters.
\item Generate $d_{\max}$ uniformly from $d_{\min}\sim70$ meters.
\item Generate D2D links distance as uniform $d_{\min}\sim d_{\max}$.
\end{itemize}

As noted earlier, we could have also used a state-of-the-art optimization
algorithm to generate locally optimal schedules as targets and train the neural
network in a supervised fashion. 
Promising results have been obtained for specific transmitter-receiver distance
distributions (e.g., 2$\sim$65 meters) \cite{Cui_GC}, but supervised learning 
does not always work well for more general distributions; see Section \ref{sec:unsupervisedVSsupervised}.
A possible explanation is that the high quality local optimal schedules are
often not a smooth functional mapping of the network parameters, and are
therefore difficult to learn. 

\subsection{Symmetry Breaking}

\begin{figure}[t]
        \centering
    	\ifOneColumn
\else
            \includegraphics[width=8cm]{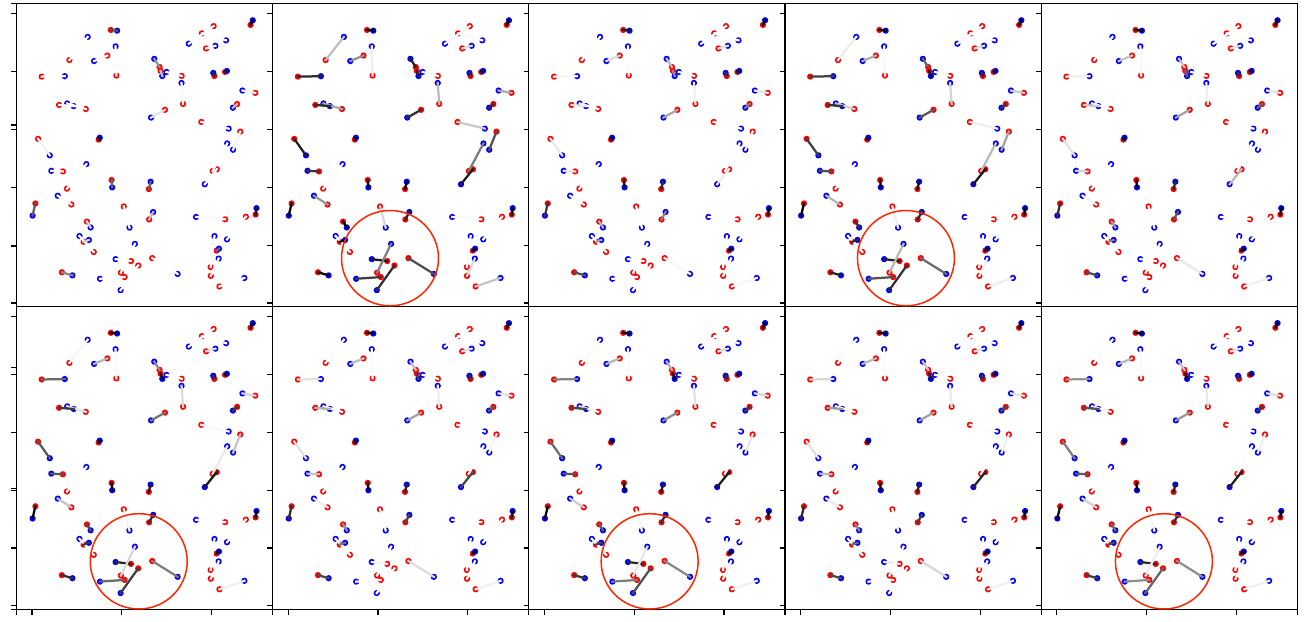}
        \fi
        \caption{Oscillatory behavior in the neural network training process.}
        \label{fig: evoSymmetry}
\end{figure}

The overall neural network is designed to encourage links to deactivate either
when it produces too much interference to its neighbors, or when it experiences
too much interference from its neighbors. However, because training happens in
stages and all the links update their activation pattern in parallel,
the algorithm frequently gets into situations in which multiple links may oscillate
between being activated and deactivated.

Consider the following scenario involving two closely located links with
identical surroundings. Starting from the initialization stage where both
links are fully activated, both links see severe interference coming from
each other. Thus, at the end of the first forward path, both links would
be turned off. Now assuming that there are no other strong
interference in the neighborhood, then at the end of the second iteration,
both links would see little interference; consequentially both would be
encouraged to be turned back on.  This oscillation pattern can keep going, and
the training process for the neural network would never converge to a good schedule
(which is that one of the two links should be on).
Fig.~\ref{fig: evoSymmetry} shows a visualization of the phenomenon.  Activation
patterns produced by the actual training process are shown in successive snapshots.
Notice that the closely located strong interfering links located at
middle bottom of the layout have the oscillating pattern between successive
iterations. The training procedure does not converge to
a reasonably good schedule, with only one of the links being scheduled. 

To resolve this problem, a stochastic update mechanism to break the symmetry is
proposed. At the end of each forward path, the output vector $\mathbf x$
contains the updated activation pattern for all the links.  However,
instead of feeding back $\mathbf x$ directly to the next iteration, we feedback
the updated entries of $\mathbf x$ with 50\% probability (and feedback the old
entries of $\mathbf x$ with 50\% probability). This symmetry breaking is used
in both the training and testing phases and is observed to benefit the overall
performance of the neural network.

\section{Performance of Sum-Rate Maximization} \label{sec:originalExperiment}

\begin{table}[t]
\caption{Design Parameters for the Spatial Deep Neural Network}
\centering
\begin{tabular}{|l|l|c|}
\hline
Parameters & \multicolumn{2}{c|} {Values}  \\
\hline
\text{Convolution Filter Size}
& \multicolumn{2}{c|}{63 cells $\times$ 63 cells} \\
\hline
Cell Size & \multicolumn{2}{c|} {5m by 5m} \\
\hline
First Hidden Layer & \multicolumn{2}{c|} { 30 units } \\
\hline
Second Hidden Layer & \multicolumn{2}{c|} { 30 units } \\
\hline
\multirow{2}{*}{Number of Iterations}
& Training & 3$\sim$20 iterations \\ \cline{2-3}
& Testing & 20 iterations \\
\hline
\end{tabular}
\label{tab:networkParameters}
\end{table}

\subsection{Testing on Layouts of Same Size as Training Samples}

We generate testing samples of random wireless D2D network layouts of the same
number of links and the same size as the training samples, except with fixed
uniform link distance distribution between some values of $d_{\min}$ and
$d_{\max}$. The channel model is adapted from the short-range outdoor model
ITU-1411 with a distance-dependent path-loss \cite{itu1411}, over 5 MHz bandwidth at 2.4 GHz
carrier frequency, and with 1.5 m antenna height and 2.5 dBi antenna gain. The
transmit power level is 40 dBm; the background noise level is -169 dBm/Hz. We
assume an SNR gap of 6 dB to Shannon capacity to account for practical
coding and modulation.

For each specific layout and each specific channel realization, the FPLinQ
algorithm \cite{shen_ISIT17} is used to generate the sum-rate maximizing
scheduling output with a maximum number of iterations of 100.  We note that although
FPLinQ guarantees monotonic convergence for the optimization over the
continuous power variables, it does not necessarily produce monotonically
increasing sum rate for scheduling. Experimentally, scheduling outputs after
100 iterations show good numerical performance. We generate 5000 layouts for
testing in this section. 

The design parameters for the neural network are summarized in the Table
\ref{tab:networkParameters}. We compare the sum rate performance achieved by
the trained neural network with each of the following benchmarks in term of
both the average and the maximum sum rate over all the testing samples: 
\begin{itemize}
	\item {\bf All Active:} Activate all the links.
	\item {\bf Random:} Schedule each link with 0.5 probability.
	\item {\bf Strongest Links First:}
		We sort all the links according to the direct channel strength, then
		schedule a fixed portion of the strongest links. The optimal
		percentage is taken as the average percentage of the active links
		in the FP target.
	\item {\bf Greedy:} Sort all the links according to the link distance, then
schedule one link at a time. Choose a link to be active only if scheduling
this link strictly increases the objective function (i.e., the sum rate).
Note that the interference at all the active links needs to be re-evaluated in each
step as soon as a new link is turned on or off.
	\item {\bf FP:} Run FPLinQ for 100 iterations.
\end{itemize}

We run experiments with the following D2D links pairwise distance distributions
in the test samples: 
\begin{itemize}
    \item Uniform in $30\sim70$ meters.
    \item Uniform in $2\sim65$ meters.
    \item Uniform in $10\sim50$ meters.
    \item All links at $30$ meters.
\end{itemize}
The distance distribution affects the optimal scheduling strategies, e.g., in
how advantageous it is for scheduling only the strongest links.  The sum rate
performance of each of the above methods are reported in Table
\ref{tab:sumratesPure}.  The performance is expressed as the percentages as
compared to FPLinQ. 

\begin{table}
\caption{Average Sum Rate as Percentage of FP}
\centering
\begin{tabular}{|c|c||c|c|c|c|}
\ifOneColumn
\else
    \hline
    \% & CSI & 30$\sim$70 & 2$\sim$65 & 10$\sim$50 & all 30\\
    \hline
    Learning & \crossmark & 92.19 & 98.36 & 98.42 & 96.90 \\
    \hline
    Greedy & \checkmark & 84.76 & 97.08 & 94.00 & 84.56 \\
    \hline
    Strongest & \crossmark & 59.66 & 82.03 & 75.41 & N/A \\
    \hline
    Random & \crossmark & 35.30 & 47.47 & 49.63 & 50.63 \\
    \hline
    All & \crossmark & 26.74 & 54.18 & 48.22 & 43.40 \\
    \hline
    FP & \checkmark & 100 & 100 & 100 & 100 \\
    \hline
\fi
\end{tabular}
\label{tab:sumratesPure}
\end{table}

As shown in Table \ref{tab:sumratesPure}, the proposed
spatial learning approach always achieves more than 92\% of the average sum rate
produced by FPLinQ for all cases presented, \emph{without explicitly knowing the channels.} 
The neural network also outperforms the greedy heuristic (which requires CSI) and outperforms other benchmarks by large margins. 

\begin{figure}
    \centering
    \ifOneColumn
\else
        \subfigure[Greedy]{\includegraphics[width=4cm]{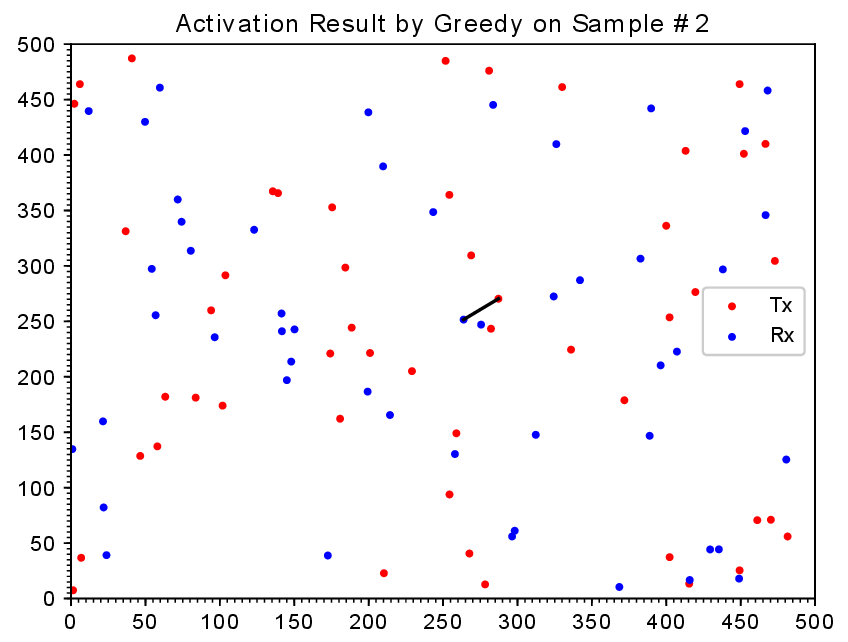}}
        \\
        \subfigure[FP]{\includegraphics[width=3.5cm]{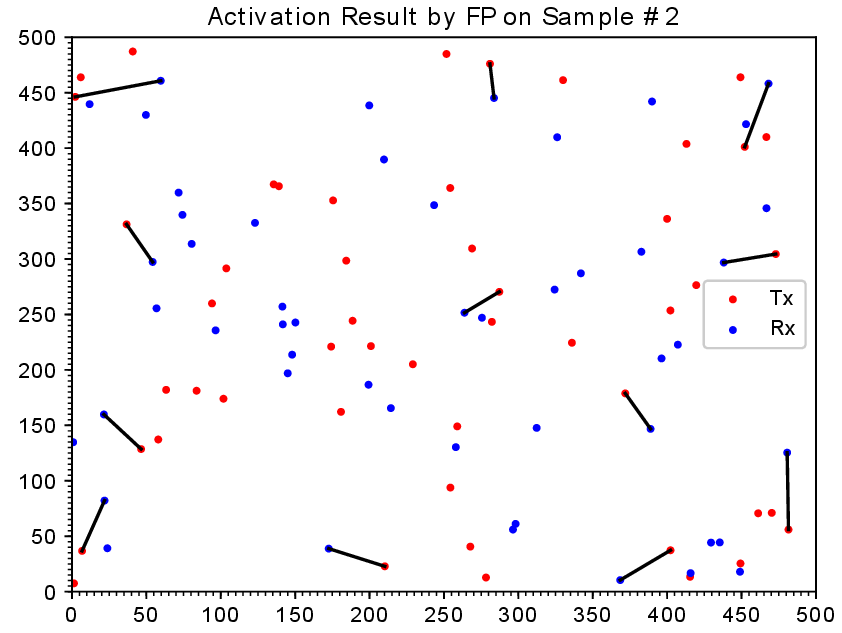}}
        \subfigure[Neural Network]{\includegraphics[width=3.5cm]{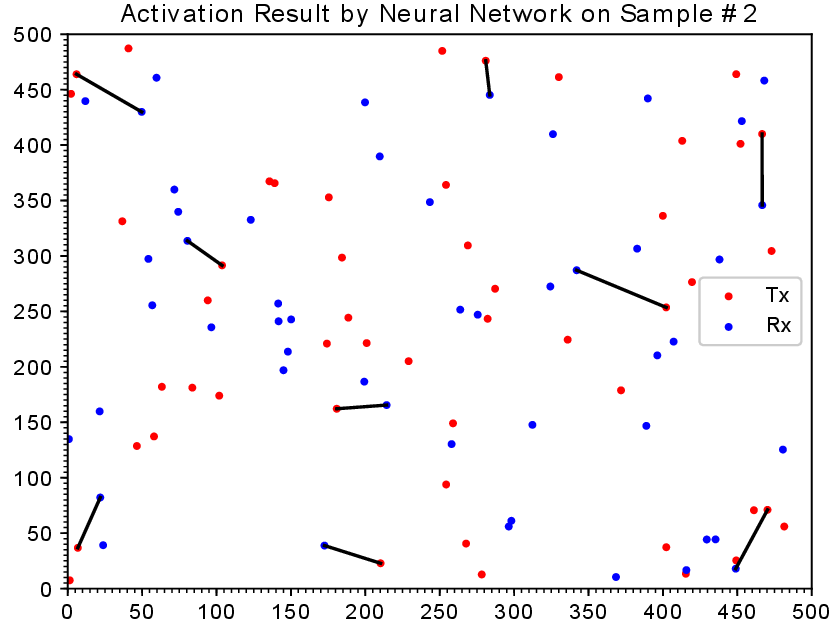}}
    \fi
    \caption{Greedy heuristic prematurely activates the strongest link}
    \label{fig:greedyprematureproblem}
\end{figure}

The main reason that the greedy heuristics performs poorly is that it always
activates the strongest link first, but once activated, the algorithm does not
reconsider the scheduling decisions already made. The earlier scheduling
decision may be suboptimal; this leads to poor performance as illustrated in
an example in Fig.~\ref{fig:greedyprematureproblem}. Note that under the
channel model used in simulation, the interference of an activated link reaches
a range of 100m to 300m. If a greedy algorithm activates a link in the center
of the 500m by 500m layout, it could preclude the activation of all other
links, while the optimal scheduling should activate multiple weaker links
roughly 100m to 300m apart as shown in Fig.~\ref{fig:greedyprematureproblem}. 

Throughout testings of many cases, including the example shown in
Fig.~\ref{fig:greedyprematureproblem}, the spatial learning approach always
produces a scheduling pattern close to the FP output. This shows that the
neural network is capable of learning the state-of-the-art optimization strategy.

\subsection{Generalizability to Arbitrary Topologies}

An important test of the usefulness of the proposed spatial deep learning
design is its ability to generalize to different layout dimensions and link
distributions.  Intuitively, the neural network performs scheduling based on 
an estimate of the direct channel and the aggregate interference from a local
region surrounding the transmitter and the receiver of each link. Since both of
these estimates are local, one would expect that the neural network should be
able to extend to general layouts.

To validate this generalization ability, we test the trained neural network on
layouts with larger number of links, while first keeping the link density the
same, then further test on layouts in which the link density is different. Note
that we do not perform any further training on the neural network. For each test, 500 random layouts are generated to obtain the average maximum sum rate. 

\subsubsection{Generalizability to Layouts of Large Sizes}
First, we keep the link density and distance distribution the same and test the
performance of the neural network on larger layouts occupying an area of
up to 2.5km by 2.5km and 1250 links. The resulting sum
rate performance is presented in Table
\ref{tab:fixDensityEval}. Note that following the earlier convention, the
entries for the deep learning (DL) neural network and greedy method are the
percentages of sum rates achieved as compared with FP, averaged over the
testing set.

\begin{table}
\caption{Generalizability to Layouts of Larger Dimensions but Same Link Density
and Link Distance: Sum Rate as \% of FP}
\centering
\ifOneColumn
\else
    \begin{tabular}{|c|c|c|c|c|c|} 
    \hline
    \multirow{2}{*}{\shortstack[c]{Size (m$^2$)}} & \multirow{2}{*}{\shortstack[c]{Links}} 
    & \multicolumn{2}{c|}{2m$\sim$65m} & \multicolumn{2}{c|}{All 30m} \\ \cline{3-6}
    & & \shortstack[c]{DL} & \shortstack[c]{Greedy} & \shortstack[c]{DL} & \shortstack[c]{Greedy} \\
    \hline
    $750 \times 750$ & 113 & 98.5 & 102.4 & 98.4 & 98.4 \\
    \hline
    $1000 \times 1000$ & 200 & 99.2 & 103.2 & 98.3 & 98.8 \\
    \hline
    $1500 \times 1500$ & 450 & 99.5 & 103.8 & 98.3 & 100.0 \\
    \hline
    $2000 \times 2000$ & 800 & 99.7 & 104.1 & 98.8 & 100.8 \\
    \hline
    $2500 \times 2500$ & 1250 & 99.7 & 104.2 & 99.1 & 101.3 \\
    \hline
    \end{tabular}
\fi
\label{tab:fixDensityEval}
\end{table}

Table \ref{tab:fixDensityEval} shows that the neural network is able to
generalize to layouts of larger dimensions very well, with performance
very close to FP. It is worth emphasizing that while the greedy algorithm also performs well
(likely because the phenomenon of Fig.~\ref{fig:greedyprematureproblem}
is less likely to occur on larger layouts), it requires CSI, as opposed to just
location information utilized by spatial deep learning.

\subsubsection{Generalizability to Layouts with Different Link Densities}
\label{sec:sumrateVaryDensity}

We further explore the neural network's generalization ability in optimizing
scheduling over layouts that have different link densities as compared to the
training set. For this part of the evaluation, we fix the layout size to be 500
meters by 500 meters as in the training set, but instead of having 50 links, we
vary the number of links in each layout from 10 to 500. The resulting sum rate 
performances of deep learning and the greedy heuristics are presented in Table
\ref{tab:varyDensityEval}.

\begin{table}
\caption{Generalizability to Layouts with Different Link Densities:
Sum Rate as \% of FP}
\centering
\ifOneColumn
\else
    \begin{tabular}{|c|c||c|c||c|c|}
    \hline
    \multirow{2}{*}{\shortstack[c]{Size ($m^2$)}} & \multirow{2}{*}{\shortstack[c]{Links}} & \multicolumn{2}{c|}{2m$\sim$65m} & \multicolumn{2}{c|}{All 30m} \\ \cline{3-6} & & DL & Greedy & DL & Greedy \\
    \hline
    \multirow{5}{*}{$500 \times 500$} & 10 & 95.5 & 90.0 & 94.9 & 81.6 \\
    \cline{2-6}
    & 30 & 97.0 & 93.2 & 96.1 & 81.3 \\
    \cline{2-6}
    & 100 & 98.6 & 99.8 & 99.0 & 88.7 \\
    \cline{2-6}
    & 200 & 97.8 & 101.7 & 96.0 & 92.4 \\
    \cline{2-6}
    & 500 & 93.0 & 104.1 & 92.9\footnotemark & 92.8 \\
    \hline
    \end{tabular}
\fi
\label{tab:varyDensityEval}

{\footnotesize $^1$
50 iterations are required for deep learning to achieve this result}
\end{table}

As shown in Table \ref{tab:varyDensityEval}, with up to 4-fold increase in 
the density of interfering links, the neural network is able to perform near
optimally, achieving almost the optimal FP sum rate, while significantly
outperforming the greedy algorithm, especially when the network is sparse. 

However, the generalizability of deep learning does have limitations. When 
the number of links increases to 500 or more (10-fold increase as compared to
training set), the neural network becomes harder to converge, resulting in
dropping in performance. This is reflected in one entry in the last row of
Table \ref{tab:varyDensityEval}, where it takes 50 iterations for the neural
network to reach a satisfactory rate performance. If the link density is
further increased, it may fail to converge. Likely, new training set with
higher link density would be needed.

\subsection{Sum Rate Optimization with Fast Fading}

So far we have tested on channels with only a path-loss component (according
to the ITU-1411 outdoor model). Since path-loss is determined by location,
the channels are essentially deterministic function of the location.

In this section, Rayleigh fast fading is introduced into the testing channels.
This is more challenging, because the channel gains are now stochastic
functions of GLI inputs. Note that the neural network is still trained using
channels without fading. 

We use test layouts of 500 meters by 500 meters with 50 D2D links and 
with 4 uniform link distance distributions. The sum rate performance results
are presented in Table \ref{tab:sumratesFastfading}, with an additional benchmark:
\begin{itemize}
    \item {\bf FP without knowing fading:} Run FP based on the CSI without
fast fading effect added in. This represents the best that one
can do without knowing the fast-fading.  
\end{itemize}

\begin{table}
\caption{Sum Rate as \% of FP on Channels with Fast Fading}
\centering
\begin{tabular}{|c|c||c|c|c|c|}
\ifOneColumn
\else
    \hline
    \% & CSI & 30$\sim$70 & 2$\sim$65 & 10$\sim$50 & 30\\
    \hline
    DL & \crossmark & 71.8 & 88.6 & 82.5 & 73.9 \\
    \hline
    FP no fade & \checkmark & 77.7 & 88.9 & 82.7 & 76.3 \\ 
    \hline
    Greedy & \checkmark & 95.9 & 98.3 & 97.7 & 96.7 \\
    \hline
    Strongest & \checkmark & 65.4 & 80.8 & 75.0 & 68.8 \\
    \hline
    Random & \crossmark & 31.7 & 44.5 & 44.0 & 42.7 \\
    \hline
    All Active & \crossmark & 25.3 & 50.4 & 43.8 & 38.4 \\
    \hline
    FP & \checkmark & 100 & 100 & 100 & 100 \\
    \hline
\fi
\end{tabular}
\label{tab:sumratesFastfading}
\end{table}

As shown in Table \ref{tab:sumratesFastfading}, the performance of deep
learning indeed drops significantly as compared to FP or Greedy (both of which
require exact CSI). However, it is still encouraging to see that the performance of
neural network matches \emph{FP without knowing fading}, indicating that it is
already performing near optimally given that only GLI is available as inputs. 

\subsection{Computational Complexity}

In this section, we further argue that the proposed neural network has a
computation complexity advantage as compared to the greedy or FP algorithms by
providing a theoretical analysis and some experimental verification. 

\subsubsection{Theoretical Analysis}
We first provide the complexity scaling for each of the methods as functions of
the number of links $N$:
\begin{itemize}
	\item \textbf{FPLinQ Algorithm}:
    	Within each iteration, to update scheduling outputs and relevant quantities, the dominant computation includes matrix multiplication with the $N\times N$ channel coefficient matrix. Therefore, the complexity per iteration is $O(N^2)$. Assuming that a constant number of iterations is needed for convergence, the total run-time complexity is then $O(N^2)$.
   \item \textbf{Greedy Heuristic}: The greedy algorithm makes scheduling decisions for each link sequentially. When deciding whether to schedule the $i$th link, it needs to compare the sum rate of all links that have been scheduled so far, with and without activating the new link. This involves re-computing the interference, which costs $O(i)$ computation. As $i$ ranges from $1$ to $N$, the overall complexity of the greedy algorithm is therefore $O(N^2)$.
	\item \textbf{Neural Network} Let the discretized grid be of dimension $K\times K$, and the spatial filter be of dimension $J\times J$. Furthermore, let $h_0$ denotes the size of input feature vector for fully connected stage, and let $(h_1, h_2,...h_n)$ denote the number of hidden units for each of the $n$ hidden layers (note that the output layer has one unit). The total run-time complexity of the proposed neural network can be computed as:
\begin{align}
\underbrace{K^2 \times J^2}_{\text{Convolution Stage}} + N \times
\underbrace{(h_0h_1 + \dots + h_{n-1}h_{n} + h_{n})}_{\text{Fully Connected Stage per Pair}}
\end{align}
Thus, given a layout of fixed region size, the time complexity of neural
network scales as $O(N)$. 
\end{itemize} 

\subsubsection{Experimental Verification}
In actual implementations, due to its ability to utilize parallel computation
architecture, the run-time of the neural network can be even less than $O(N)$. 
To illustrate this point, we measure the total computation time of scheduling one layout of varying number of D2D
links by using FP and greedy algorithms and by using the proposed neural
network.  The timing is conducted on a single desktop, with the hardware
specifications as below:
\begin{itemize}
    \item FP and Greedy: Intel CPU Core i7-8700K @ 3.70GHz
    \item Neural Network: Nvidia GPU GeForce GTX 1080Ti
\end{itemize}
To provide reasonable comparison of the running time, we select hardwares most
suited for each of the algorithms. The implementation of the neural network is
highly parallelizable; it greatly benefits from the parallel computation power
of GPU. On the other hand, FP and greedy algorithms have strictly sequential
computation flows, thus benefiting more from CPU due to its much higher clock
speed. The CPU and GPU listed above are selected at about the same level of
computation power and price point with regard to their respective classes.

\begin{figure}
        \ifOneColumn
\else
            \includegraphics[width=8.5cm,height=6cm]{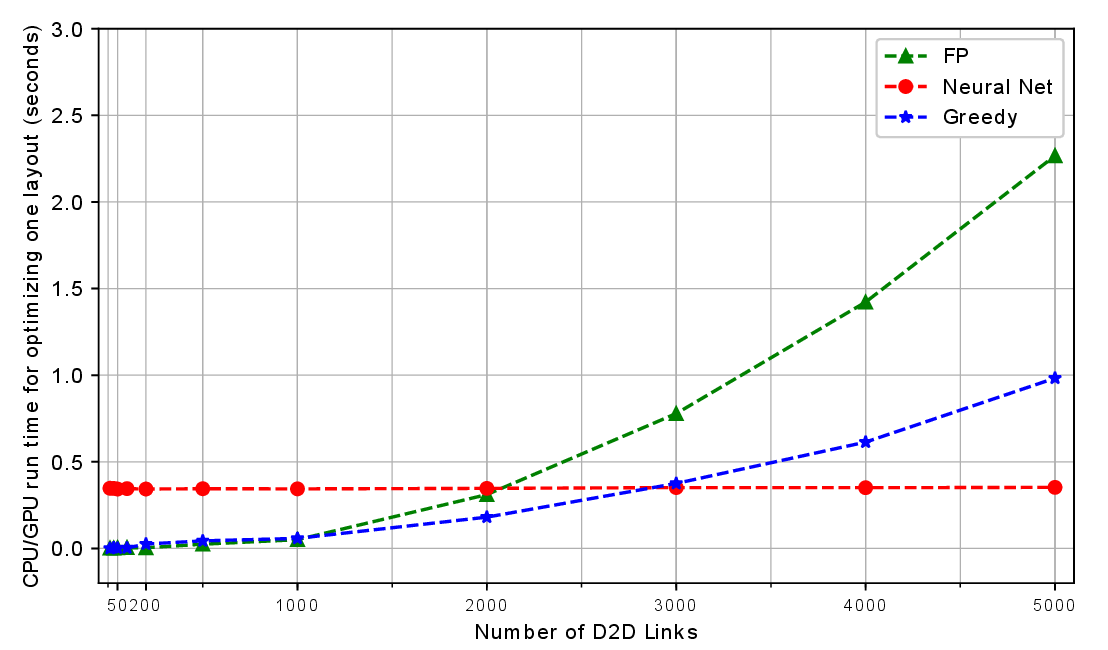}
        \fi
        \caption{Computation time on layouts with varying number of D2D links.}
        \label{fig:varyDensityTime}
\end{figure}

As illustrated in Fig.~\ref{fig:varyDensityTime}, the computational complexity
of the proposed neural network is approximately constant, and is indeed several
orders of magnitude less than FP baseline for layouts with large number of D2D
links.  

We remark here that the complexity comparison is inherently implementation
dependent. For example, the bottleneck of our neural network implementation is
the spatial convolutions, which are computed by built-in functions in TensorFlow
\cite{tensorflow}. The built-in function for computing 
convolution in TensorFlow, however, computes convolution in every location in
the entire geographic area, which is an overkill. If a customized convolution
operator is used only at specific locations of interests, the rum-time complexity
of our neural network can be further reduced. The complexity is expected to be
$O(N)$, but with much smaller constant than the complexity curve in
Fig.~\ref{fig:varyDensityTime}. We also remark that the computational
complexity of traditional optimization approaches can potentially be reduced by
further heuristics; see, e.g., \cite{Guo_1000}.

To conclude, the proposed neural network has significant computational
complexity advantage in large networks, while maintaining near-optimal
scheduling performance. This is remarkable considering that the neural network
has only been trained on layouts with 50 links, and requires only $O(N)$ GLI
rather than $O(N^2)$ CSI.

\subsection{Unsupervised vs. Supervised Training} \label{sec:unsupervisedVSsupervised}

As mentioned earlier, the neural network can also be trained in a supervised 
fashion using the locally optimal schedule from FP, or in a unsupervised fashion
directly using the sum-rate objective. Table \ref{tab:supervisedVSunsupervised}
compares these two approaches on the layouts of 500 meters by 500 meters with
50 D2D links, but with link distances following four different distributions. 
It is interesting to observe that while supervised learning is competitive for
link distance distribution of 2m$\sim$65m, it generally has inferior performance
in other cases. An intuitive reason is that when the layouts contain very short
links, the sum-rate maximization scheduling almost always chooses these short
links. It is easier for the neural network to learn such pattern in
either a supervised or unsupervised fashion. When the layouts contain links of
similar distances, many distinct local optima emerge, which tend to confuse the
supervised learning process. In these cases, using the sum-rate objective
directly tends to produce better results. 

\begin{table}
\caption{Unsupervised vs. Supervised Training -- Sum Rate as \% of FP}
\centering
\begin{tabular}{|c||c|c|c|c|}
\hline
Sum Rate (\%) & 2$\sim$65m & 10$\sim$50m & 30$\sim$70m & 30m\\
\hline
Unsupervised & 98.4 & 98.4 & 92.2 & 96.9 \\
\hline
Supervised & 96.2 & 90.3 & 83.2 & 82.0 \\ 
\hline
\end{tabular}
\label{tab:supervisedVSunsupervised}
\end{table}


\section{Scheduling with Proportional Fairness}

This paper has thus far focused on scheduling with sum-rate objective, which
does not include a fairness criterion, thus tends to favor shorter links and links
that do not experience large amount of interference. Practical applications
of scheduling, on the other hand, almost always require fairness.  In the
remaining part of this paper, we first illustrate the challenges in incorporating
fairness in spatial deep learning, then offer a solution that takes advantage
of the existing sum-rate maximization framework to provide fair scheduling
across the network.

\subsection{Proportional Fairness Scheduling}

We can ensure fairness in link scheduling by defining an optimization
objective of a network utility function over the long-term average
rates achieved by the D2D links. The long-term average rate, for example, can
be defined over a duration of $T$ time slots, with an exponential weighted window:
\begin{equation}
	\bar R_i^t = (1-\alpha) \bar R_i^{t-1} + \alpha R_i^t \quad t\leq T
\end{equation}
where $R_i^t$ is the instantaneous rate achieved by the D2D link $i$ in
time slot $t$, which can be computed as in (\ref{equ:instantRate}) based on
the scheduling decision binary vector in each time slot, $\mathbf{x}^t$.
Define a concave and non-decreasing utility function $U(\bar{R_i})$ for each
link. The network utility maximization problem is that of maximizing
\begin{equation} \label{equ:utility}
\sum_{i=1}^{N} U(\bar R_i).
\end{equation}
In the proportional fairness scheduling, the utility function is chosen to be
$U(\cdot) = \log(\cdot)$.

The idea of proportional fairness scheduling is to maximize the quantity
defined in (\ref{equ:utility}) \emph{incrementally} \cite{tse_pf}.
Assuming large $T$, in each new time slot, the incremental contribution of the
achievable rates of the scheduled links to the network utility is approximately
equivalent to a weighted sum rate \cite{berry}
\begin{equation}\label{equ:weightedSumRate}
	\sum_{i=1}^N w_i R_i^t
\end{equation}
where the weights are set as:
\begin{equation} \label{equ:computeWeight}
	w_i = \left.\frac{\partial U(\bar R_i^t)}{\partial R}\right|_{\bar R_i^t}
    	= \left.\frac{\partial \log(\bar R_i^t)}{\partial R}\right|_{\bar R_i^t}
        = \frac{1}{\bar R_i^t}.
\end{equation}
Thus, the original network utility
maximization problem (\ref{equ:utility}) can be solved by a series of weighted
sum-rate maximization, where the weights are updated in each time slot as in
(\ref{equ:computeWeight}).  The approximate mathematical equivalence of (\ref{equ:utility}) to this series of weighted sum-rate maximization
(\ref{equ:weightedSumRate})-(\ref{equ:computeWeight}) is established in \cite{kushner}.
In the rest of the paper, to differentiate the weights in the weighted rate-sum
maximization from the weights in the neural network, we refer $w_i$ as 
the \emph{proportional fairness weights}.

The weights $w_i$ can take on any positive real values. This presents a
significant challenge to deep learning based scheduling. In theory, one could
train a different neural network for each set of weights, but the complexity of
doing so would be prohibitive. To incorporate $w_i$ as an extra input to the neural
network turns out to be quite difficult as well. We explain this point in the
next section, then offer a solution.

\subsection{Challenge in Learning to Maximize Weighted Sum Rate}

A natural idea is to incorporate the proportional fairness weights as an extra
input for each link in the neural network. However, this turns out to be quite
challenging. We have implemented both the spatial convolution based neural network (using
the structure mentioned in the first part of the paper, while taking an extra
proportional fairness weight parameter) and the most general fully connected neural
network to learn the mapping from the proportional fairness weights to the
optimal scheduling. With millions of training data, the neural network is
unable to learn such a mapping, even for a single fixed layout. 

\begin{figure}
\centering
\ifOneColumn
\else
    \centerline{\includegraphics[width=10cm,height=4cm]{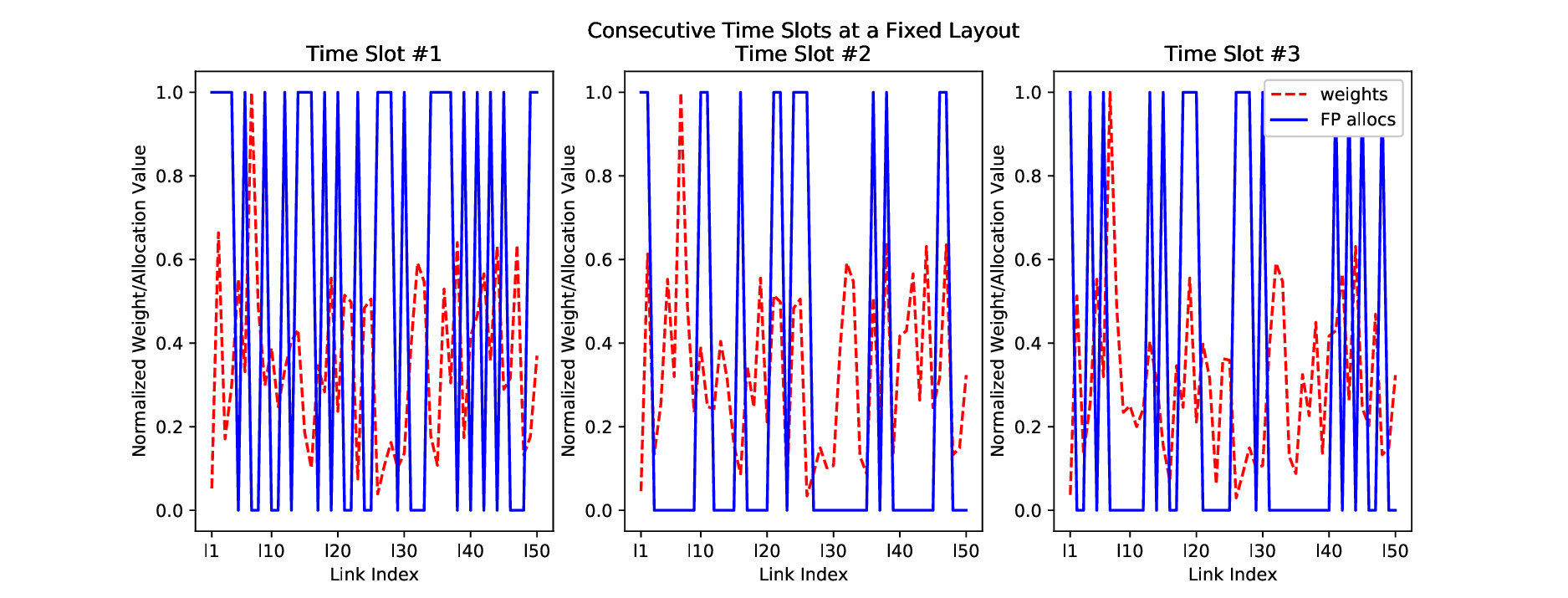}}
\fi
\caption{The optimal scheduling can drastically change over slow varying
proportional fairness weights}
\label{fig:weights_allocs}
\end{figure}

The essential difficulty lies in the high dimensionality of the function mapping. 
To visualize this complexity, we provide a series of plots of proportional
fairness weights against FP scheduling allocations in sequential time slots in
Fig.~\ref{fig:weights_allocs}. It can be observed that the FP schedule can change drastically when the proportional weights only vary by a small amount. This is indeed a feature of proportional fairness
scheduling: an unscheduled link sees its average rate decreasing and its
proportional fairness weight increasing over time until they cross a threshold,
then all the sudden it gets scheduled. Thus, the mapping between the
proportional fairness weights and the optimal schedule is highly sensitive to
these sharp turns. If we desire to learn this mapping from a data-driven approach, one should
expect to need a considerably larger amount of training samples to be collected
just to be able to survey the functional landscape, not to mention the many
more local sharp optima that would make training difficult. Further exacerbating
the difficulty is the fact that there is no easy way to sample the space of
proportional fairness weights. In a typical scheduling process, the sequence of
weights are highly non-uniform.

\subsection{Weighted Sum Rate Maximization via Binary Weights} \label{sec:stepByStepSumRate}

To tackle the proportionally fair scheduling problem, this paper proposes
the following new idea. Since the neural network proposed in the first
part of this paper is capable of generalizing to arbitrary topologies for
sum-rate maximization, we take advantage of this ability by emulating weighted
sum-rate maximization by sum-rate maximization, but over a judiciously chosen
subset of links.

The essence of scheduling is to select an appropriate subset of users to activate.
Our idea is therefore to first construct a shortlist of candidate links based on the
proportional fairness weights alone, then further refine the candidate set of
links using deep learning. Alternatively, this can also be thought of as to
approximate the proportional fairness weights by a binary weight vector 
taking only the values of 0 or 1.

The key question is how to select this initial shortlist of candidate links,
or equivalently how to construct the binary weight vector.  Denote the original
proportional fairness weights as described in (\ref{equ:computeWeight}) by
$\mathbf{w^t}$. Obviously, the links with higher weights should have higher priority. The
question is how many of the links with the large weights should be included.

This paper proposes to include the following subset of links. We think of the
problem as to approximate $\mathbf{w^t}$ by a binary 0-1 vector
$\mathbf{\hat{w}^t}$.  The proposed scheme finds this binary approximation in
such a way so that the dot product between $\mathbf{w^t}$ (normalized to 
unit $\ell_2$-norm) and $\mathbf{\hat{w}^t}$ (also normalized) is maximized.
For a fixed real-valued weight vector $\mathbf{w^t}$, we find the binary
weight vector $\mathbf{\hat{w}^t}$ as follows:
\begin{equation}
\mathbf{\hat{w}^t} = \displaystyle \mathrm{arg}\max_{\mathbf{y} \in \{0,1\}^N} 
\left\langle \frac{\mathbf{y}}{\|\mathbf{y}\|_2} \ , \
\frac{\mathbf{w^t}}{\|\mathbf{w^t}\|_2}  \right\rangle 
\label{binary_weights}
\end{equation}
where $\langle \cdot , \cdot \rangle$ denotes the dot product of two vectors.
Geometrically, this amounts to finding an $\mathbf{\hat{w}^t}$ that is
closest to $\mathbf{w^t}$ in term of the angle between the two vectors.  

Algorithmically, such a binary vector can be easily found by
first sorting the entries of $\mathbf{w^t}$, then setting the largest $k$
entries to 1 and the rest of 0, where $k$ is found by a linear search using the
objective function in (\ref{binary_weights}). 
With the binary weight vector $\mathbf{\hat{w}^t}$, the weighted sum rate
optimization is reduced to sum rate optimization, over the subset
of links with weights equal to 1. We can then utilize spatial deep learning to perform
scheduling over this subset of links. 

\subsection{Utility Analysis of Binary Reweighting Scheme}
\label{sec:binaryWeightsAnalysis}

The proposed binary reweighting scheme is a heuristic for producing fair user scheduling, but a rigorous analysis of such a scheme is challenging. In the following, we provide a justification as to why such a scheme provides fairness. From a stochastic
approximation perspective \cite{kushner}, the proposed way of updating the weights can be thought of as maximizing a particular utility function of the long-term
average user rate. To see what this utility function looks like, we start with
a simple fixed-threshold scheme:
\begin{equation}
\label{w_to_binary}
	\hat w_i = \left\{
\begin{aligned}
&1,\;\text{if}\;w_i\ge\theta\\
&0,\;\text{otherwise}
\end{aligned} \right.
\end{equation}
for some fixed threshold $\theta>0$, where $\hat w_i$ and $w_i$ are the binary
weight and the original weight, respectively. Since $w_i=1/\bar R_i$, we can
rewrite (\ref{w_to_binary}) as
\begin{equation}
\label{R_to_binary}
	\hat w_i = \left\{
\begin{aligned}
&1,\;\text{if}\;\bar R_i\le\frac{1}{\theta}\\
&0,\;\text{otherwise}
\end{aligned} \right..
\end{equation}
Recognizing (\ref{R_to_binary}) as a reverse step function with sharp transition
from 1 to 0 at $1/\theta$, we propose to use the following \emph{reverse sigmoid}
function to mimic $\hat w_i$:
\begin{equation}
\label{sigmoid_weight}
W(\bar R_i) = \frac{1}{1+\exp(\kappa(\bar R_i-\theta))}
\end{equation}
where the parameter $\kappa>0$ controls the steepness of the $W(\bar R_i)$. We can now recover the utility function that the reweighting scheme (\ref{sigmoid_weight})
implicitly maximizes.

\begin{figure}
        \centering
        \includegraphics[width=0.5\textwidth]{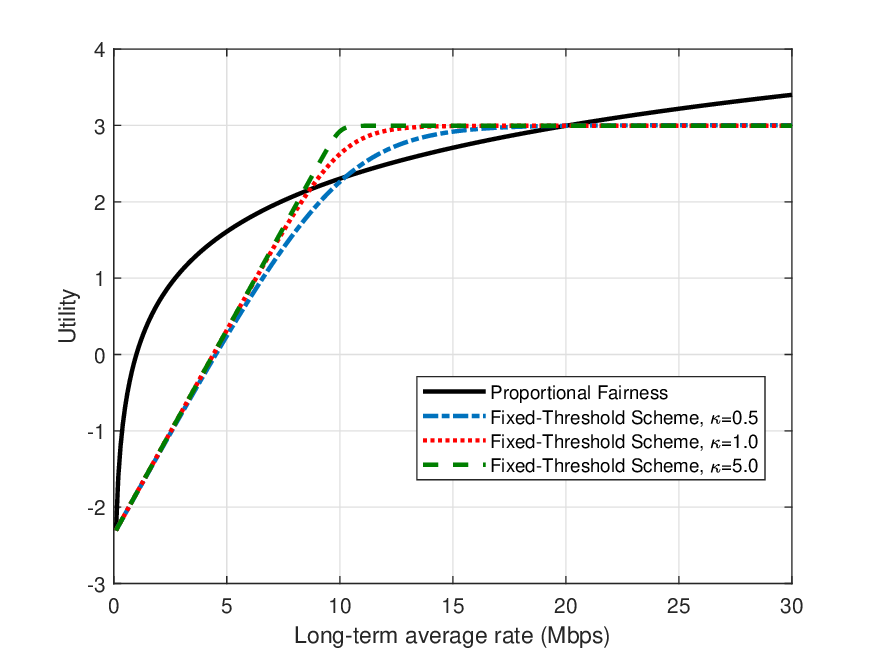}
        \caption{Utility function of the fixed-threshold binary weighting scheme 
	    vs. proportional fairness scheduling. Here $\theta=0.1$. } 
        \label{fig:new_utility}
\end{figure}

For a fixed {strictly concave} utility $U(\bar R_i)$, the user weights are set
as $w_i=U'(\bar R_i)$.  Thus, given some reweighting scheme $w_i=U'(\bar R_i)$,
the corresponding utility objective must be $U(\bar R_i)$. 
In our case, the utility function $U(\bar R_i)$ can be computed explicitly as
\begin{align}
\label{binary_utility}
U(\bar R_i) &= \alpha\int W(\bar R_i)\;d\bar R_i\notag\\
&= \alpha\bar R_i - \frac{\alpha}{\kappa}\ln\left(1+\exp(\kappa(\bar R_i-\theta))\right)+\beta
\end{align}
where $\alpha>0$ is a scaling parameter and $\beta\in\mathbb R$ is an offset
parameter. These two parameters do not affect the scheduling performance.
Fig.~\ref{fig:new_utility} compares the utility function $U(\bar R_i)$ of the
binary weighting scheme with the log-utility proportional fairness function. It
is observed that the utility of the fixed-threshold scheme follows the
same trend as the proportional fairness utility.

Note that the above simplified analysis assumes that the threshold $\theta$ is
fixed, but in the proposed binary reweighting scheme, the threshold changes
adaptively in each step, so this analysis is an approximation. Observe also that the utility function of the binary reweighting scheme
saturates when $\bar R$ is greater than the threshold, in contrast to the proportional
fairness utility which grows logarithmically with $\bar R$. This difference
becomes important in the numerical evaluation of the proposed scheme.

\section{Performance of Proportional Fairness Scheduling}
\label{sec:logutileval}

We now evaluate the performance of the deep learning based approach with binary
reweighting for proportional fairness scheduling in three types of wireless
network layouts: 
\begin{itemize}
\item The layouts with the same size and link density;
\item The larger layouts but with same link density;
\item The larger layouts but with different link density.
\end{itemize}
For testing on layouts with the same setting, 20 distinct layouts are generated for
testing, with each layout being scheduled over 500 time slots.  For the other
two settings, 10 distinct layouts are generated and scheduled over 500 time
slots. Since scheduling is performed here within a finite number of time slots,
we compute the mean rate of each link by averaging the instantaneous
rates over all the time slots:
\begin{equation} \label{equ:simpleMeanRate}
	\bar R_{i} = \frac{1}{T}\sum_{t=1}^T R_i^t.
\end{equation}
The utility of each link is computed as the logarithm of the
mean rates in Mbps. The network utility is the sum of link utilities as defined in
(\ref{equ:utility}). The utilities of distinct layouts are averaged and
presented below.  To further illustrate the mean rate distribution of the D2D
link, we also plot the cumulative distribution function (CDF) of the mean link
rates, serving as a visual illustration of fairness.

\begin{figure}
\centering
\includegraphics[width=8.5cm]{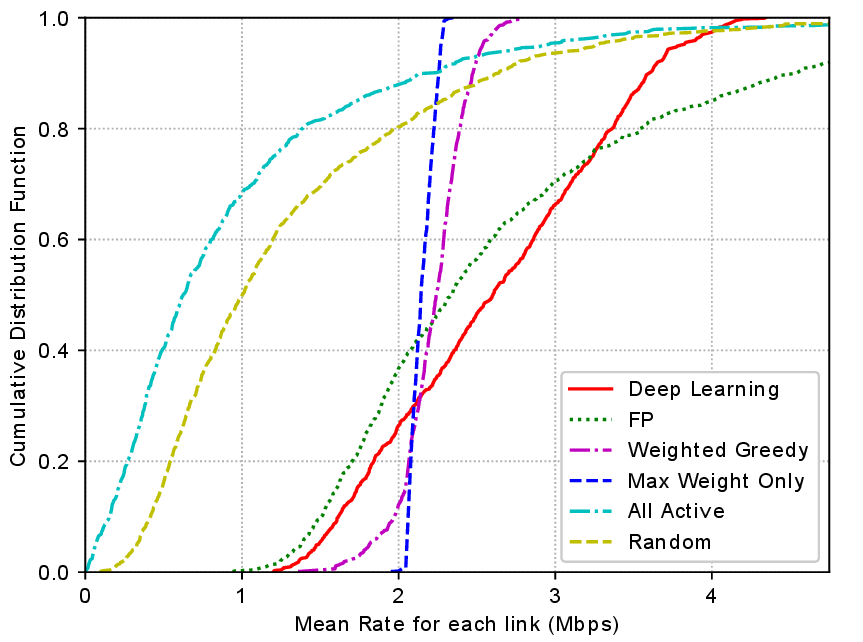}
\caption{CDF of mean rates for layouts of 50 links in 500m$\times$500m area for the
case that link distance distribution is 30m to 70m.}
\label{fig:origmeanratescdf}
\end{figure}

\begin{table}
\caption{Mean Log Utility Performance for Proportionally Fair Scheduling}
\centering
\begin{tabular}{|c|c||c|c|c|c|}
\ifOneColumn
\else
    \hline
     & CSI & 30-70 & 2-65 & 10-50 & 30\\
    \hline
    DL & \crossmark & 45.9 & 61.9 & 63.3 & 62.6 \\
    \hline
    W. Greedy & \checkmark & 39.7 & 51.5 & 51.1 & 49.0 \\
    \hline
    Max Weight & \crossmark & 38.3 & 42.1 & 41.9 & 41.4 \\
    \hline
    Random & \crossmark & 0.76 & 38.4 & 38.7 & 35.1 \\
    \hline
    All Active & \crossmark & -27.6 & 24.0 & 20.9 & 15.7 \\
    \hline
    FP & \checkmark & 45.2 & 63.1 & 63.3 & 63.0 \\
    \hline
\fi
\end{tabular}
\label{tab:propOrig}
\end{table}

The proposed deep learning based proportional fairness scheduling solves a
sum-rate maximization problem over a subset of links using the binary
reweighting scheme in each time slot. In addition to the baseline schemes
mentioned previously, we also include:
\begin{itemize}
\item {\bf Max Weight:} Schedule the single link with the highest proportional fairness weight in each time slot.
\item {\bf Weighted Greedy:} Generate a fixed ordering of all links by
sorting all the links according to the proportional fairness weight of each
link multiplied by the maximum direct link rate it can achieve without
interferences, then schedule one link at a time in this order. Choose a
link to be active only if scheduling this link strictly increases the weighted
sum rate. Note that interference is taken into account when computing the link
rate in the weighted sum rate computation. Thus, CSI is required. In fact, the
interference at all active links needs to be re-evaluated in each step whenever
a new link is activated.
\end{itemize}

\subsubsection{Performance on Layouts of Same Size and Link Density}

In this first case, we generate testing layouts with size 500
meters by 500 meters, with 50 D2D links in each layout. Similar to sum rate optimization evaluation, we have conducted the testing under the following 4 D2D links pairwise distance distributions: 
\begin{itemize}
    \item Uniform in $30\sim70$ meters.
    \item Uniform in $2\sim65$ meters.
    \item Uniform in $10\sim50$ meters.
    \item All $30$ meters.
\end{itemize}
The log utility values achieved by the various schemes are presented in Table
\ref{tab:propOrig}.  The CDF plot of mean rates achieved for the case of link
distributed in 30m-70m is presented in Fig.~\ref{fig:origmeanratescdf}.

Remarkably, despite the many approximations, the deep learning approach with
binary reweighting achieves excellent log-utility values as compared to the FP.
Its log-utility also exceeds the weighted greedy algorithm noticeably. We again
emphasize that this is achieved with geographic information only without
explicit CSI.

It is interesting to observe that the deep learning approach has a better CDF
performance as compared to the FP in the low-rate regime, but worse mean rate
beyond the 80-percentile range. This is a consequence of the fact that the
implicit network utility function of the binary reweighting scheme is higher
than proportional fairness utility at low rates, but saturates at high rate, as
shown in Fig.~\ref{fig:new_utility}.

\begin{table*}
\caption{Mean Log Utility Performance for Proportionally Fair Scheduling on
Larger Layouts with Same Link Density}
\centering
\ifOneColumn
\else
    \begin{tabular}{|c|c||c|c|c|c|c|c|}
    \hline
    \multirow{2}{*}{\shortstack[c]{Layout Size}} & \multirow{2}{*}{\shortstack[c]{Links}} 
    & \multicolumn{3}{c|}{\shortstack[c]{2m$\sim$65m}} 
    & \multicolumn{3}{c|}{\shortstack[c]{all 30 m}} \\ \cline{3-8}
    & & FP & DL & W. Greedy & FP & DL & W. Greedy \\
    \hline
    $750$m $\times 750$m & 113 & 127 & 124 & 106 & 127 & 126 & 111 \\
    \hline
    $1000$m $\times 1000$m & 200 & 217 & 205 & 203 & 219 & 214 & 205 \\
    \hline 
    $1500$m $\times 1500$m & 450 & 462 & 432 & 454 & 466 & 448 & 462 \\
    \hline
    \end{tabular}
\fi
\label{tab:propSameDensity}
\end{table*}

\subsubsection{Performance on Larger Layouts with Same Link Density}

To demonstrate the ability of the neural network to generalize to layouts of
larger size under the proportional fairness criterion, we conduct further
testing on larger layouts with the same link density. 
We again emphasize that no further training is conducted. We test 
the following two D2D links pairwise distance distributions:
\begin{itemize}
    \item Uniform in $2\sim65$ meters.
    \item All $30$ meters.
\end{itemize}
The results for this setting are summarized in Table \ref{tab:propSameDensity}.

It is observed that under the proportional fairness criterion, the spatial deep
learning approach still generalizes really well. It is competitive with respect
to both FP and the weighted greedy methods, using only $O(N)$ GLI as input and 
using the binary weight approximation.

\begin{table*}[t]
\caption{Mean Log Utility Performance for Proportionally Fair Scheduling on Layouts with Different Link Density}
\centering
\ifOneColumn
\else
    \begin{tabular}{|c|c||c|c|c|c|c|c|}
    \hline
    \multirow{2}{*}{\shortstack[c]{Layout Size}} & \multirow{2}{*}{\shortstack[c]{Links}} 
    & \multicolumn{3}{c|}{\shortstack[c]{2m$\sim$65m}} 
    & \multicolumn{3}{c|}{\shortstack[c]{all 30m}} \\ \cline{3-8}
    & & FP & DL & W. Greedy & FP & DL & W. Greedy \\
    \hline
    \multirow{3}{*}{$500$m $\times 500$m} & 30 & 52 & 49 & 47 & 50 & 50 & 44 \\
    \cline{2-8}
     & 200 & -11 & -13 & -90 & -11 & -26 & -102 \\
    \cline{2-8}
     & 500 & -511 & -514 & -736 & -485 & -542 & -739 \\
    \hline
    \end{tabular}
\fi
\label{tab:propDifferentDensity}
\end{table*}

\begin{figure}
    \centering
    \includegraphics[width=8.6cm]{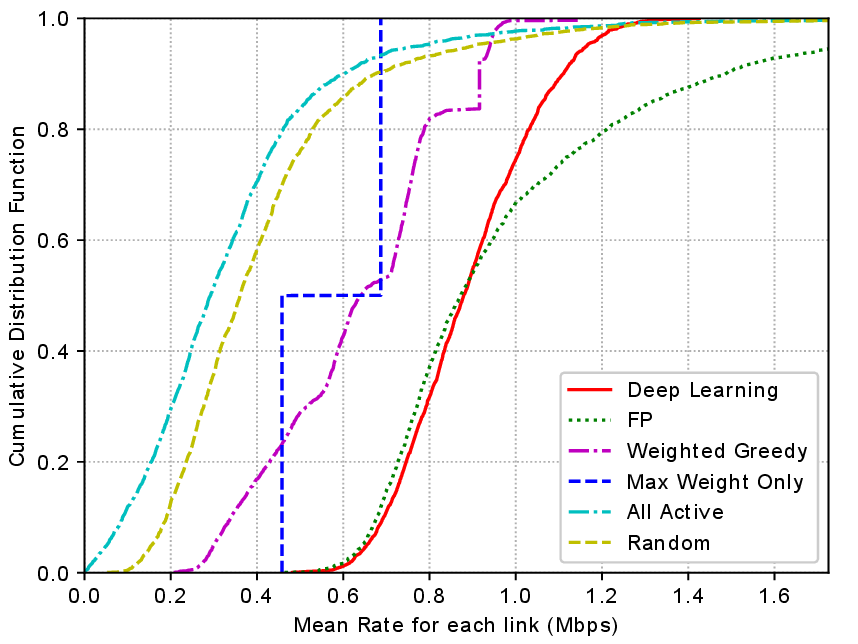}
    \caption{CDF for mean rates of Layouts with 200 links in 500m$\times$500m
area with link distance fixed at 30m}
    \label{fig:meanratescdf_varydensity}
\end{figure}

\subsubsection{Performance on Layout with Different Link Density}

We further test the neural network on a more challenging case: layouts with
different link densities than the setting on which it is trained.
Specifically, we experiment on layouts of 500 meters by 500 meters size and
varying number of D2D links. The resulting sum log utility value, averaged
over 10 testing layouts, are summarized in Table \ref{tab:propDifferentDensity}.

It is observed that the neural network still competes really well against FP in
log utility, and outperforms the weighted greedy method significantly.
To visualize, we select one specific layout of 500 meters by 500 meters
region with 200 links with link distances fixed to 30 meters, and provide
the CDF plot of long-term mean rates achieved by each link in
Fig.~\ref{fig:meanratescdf_varydensity}.

\section{Conclusion}

Deep neural network has had remarkable success in many machine learning tasks,
but the ability of deep neural networks to learn the outcome of large-scale
discrete optimization in still an open research question. This paper
provides evidence that for the challenging scheduling task for the wireless D2D
networks, deep learning can perform very well for sum-rate maximization.  
In particular, this paper demonstrates that in certain network environments, 
by using a novel geographic spatial
convolution for estimating the density of the interfering neighbors around each
link and a feedback structure for progressively adjusting the link activity
patterns, a deep neural network can in effect learn the network interference
topology and perform scheduling to near optimum based on the geographic spatial
information alone, thereby eliminating the costly channel estimation stage.

Furthermore, this paper demonstrates the generalization ability of the neural
network to larger layouts and to layouts of different link density (without the
need for any further training).  This ability to generalize provides
computational complexity advantage for the neural network on larger wireless
networks as compared to the traditional optimization algorithms and the
competing heuristics.

Moreover, this paper proposes a binary reweighting scheme to allow the weighted
sum-rate maximization problem under the proportional fairness scheduling
criterion to be solved using the neural network. The proposed method achieves
near optimal network utility, while maintaining the advantage of bypassing the
need for CSI.

Taken together, this paper shows that deep learning is promising for wireless
network optimization tasks, especially when the models are difficult or
expensive to obtain and when computational complexity of existing approaches 
is high. In these scenarios, a carefully crafted neural network topology
specifically designed to match the problem structure can be competitive to the
state-of-the-art methods.

\bibliographystyle{IEEEtran}
\bibliography{IEEEabrv,learning}

\begin{IEEEbiography}
[{\includegraphics[width=1in,height=1.25in,clip,keepaspectratio]{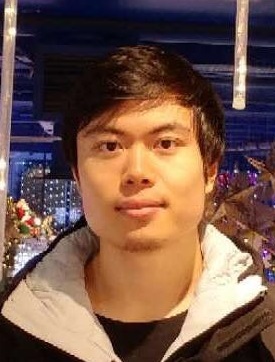}}]
{Wei Cui}
(S'17) received the B.A.Sc in Engineering Science degree from University of Toronto, Toronto, Canada in 2017, and the M.A.Sc degree in Electrical and Computer Engineering from University of Toronto, Toronto, Canada in 2019. He is currently pursuing the Ph.D. degree at the University of Toronto.

His research interests include optimization, machine learning, and wireless communication.
\end{IEEEbiography}

\begin{IEEEbiography}
[{\includegraphics[width=1in,height=1.25in,clip,keepaspectratio]{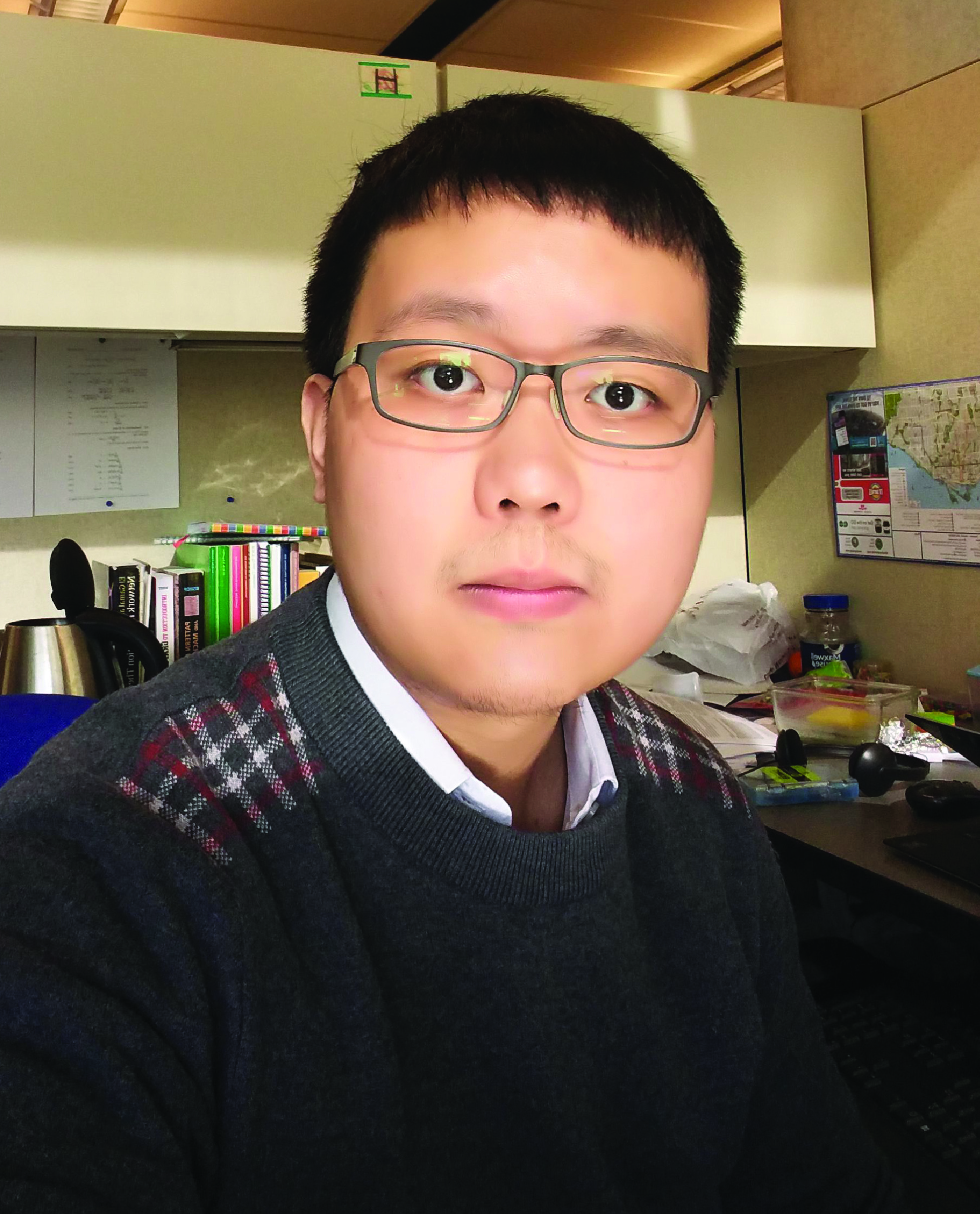}}]
{Kaiming Shen}
(S'13) received the B.Eng. degree in information security and the B.S. degree in mathematics from Shanghai Jiao Tong University, Shanghai, China in 2011, and the M.A.Sc. degree in electrical and computer engineering from the University of Toronto, Ontario, Canada in 2013. He is currently pursuing the Ph.D. degree at the University of Toronto.

His research interests include optimization, information theory, and artificial intelligence.
\end{IEEEbiography}

\begin{IEEEbiography}
[{\includegraphics[width=1in,height=1.25in,clip,keepaspectratio]{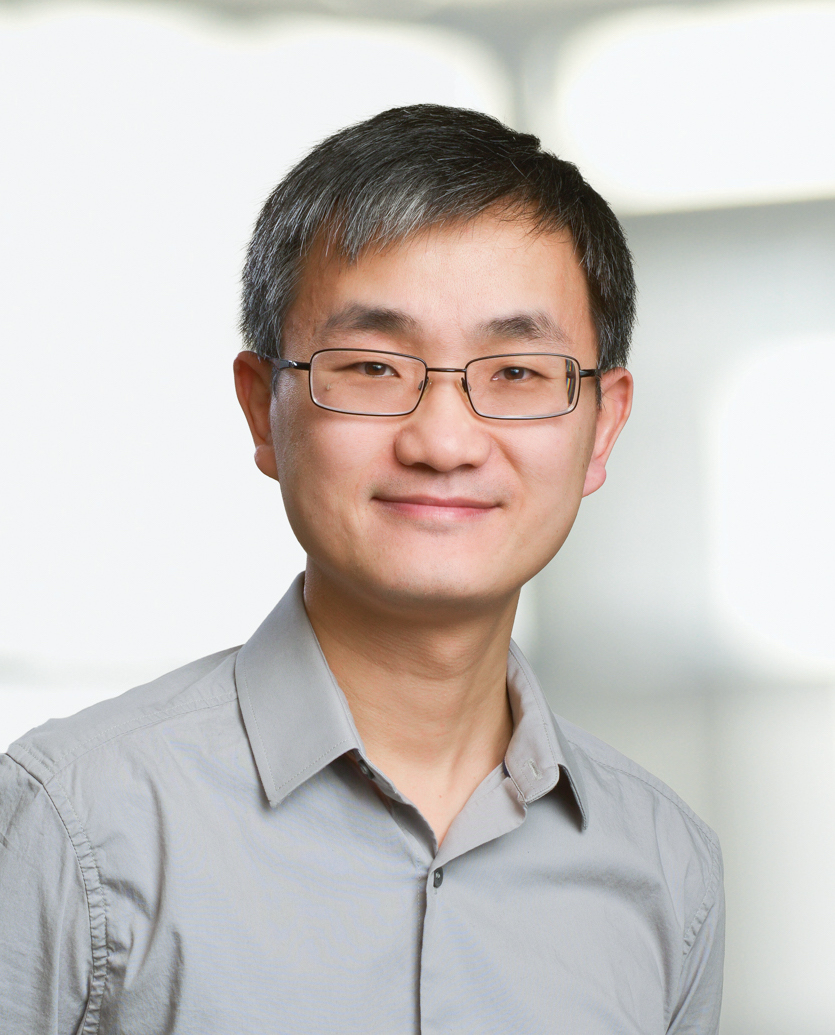}}]
{Wei Yu}
(S'97-M'02-SM'08-F’14) received the B.A.Sc. degree in Computer
Engineering and Mathematics from the University of Waterloo, Waterloo,
Ontario, Canada in 1997 and M.S. and Ph.D. degrees in Electrical
Engineering from Stanford University, Stanford, CA, in 1998 and 2002,
respectively. Since 2002, he has been with the Electrical and Computer
Engineering Department at the University of Toronto, Toronto, Ontario,
Canada, where he is now Professor and holds a Canada Research Chair
(Tier 1) in Information Theory and Wireless Communications. His main
research interests include information theory, optimization, wireless
communications, and broadband access networks.

Prof. Wei Yu serves as a Vice President of the IEEE Information Theory 
Society in 2019.
He is currently an Area Editor for the IEEE Transactions on Wireless 
Communications (2017-20),
and in the past served as an Associate Editor for IEEE Transactions on 
Information Theory
(2010-2013), as an Editor for IEEE Transactions on Communications
(2009-2011), and as an Editor for IEEE Transactions on Wireless
Communications (2004-2007). He served as the Chair of the Signal
Processing for Communications and Networking Technical Committee of the
IEEE Signal Processing Society (2017-18) and as a member in
2008-2013.
Prof. Wei Yu was an IEEE Communications Society Distinguished Lecturer 
in 2015-16.
He received the Steacie Memorial Fellowship in
2015, the IEEE Signal Processing Society Best Paper Award in 2017 and
2008, an Journal of Communications and Networks Best Paper Award in
2017, an IEEE Communications Society Best Tutorial Paper Award in 2015,
an IEEE ICC Best Paper Award in 2013, the McCharles Prize for Early
Career Research Distinction in 2008, the Early Career Teaching Award
from the Faculty of Applied Science and Engineering, University of
Toronto in 2007, and an Early Researcher Award from Ontario in 2006.
Prof. Wei Yu is a Fellow of the Canadian Academy of Engineering, and a
member of the College of New Scholars, Artists and Scientists of the
Royal Society of Canada. He is recognized as a Highly Cited Researcher.
\end{IEEEbiography}

\end{document}